\documentclass[aps,prx,floatfix,amsmath,amssymb,preprint,eqsecnum,nofootinbib,superscriptaddress, toc]{revtex4}

\usepackage{graphicx, bm, amsmath, amsfonts,amssymb,amsthm}

\usepackage{float}
\usepackage{caption}
\usepackage{color}
\usepackage{ulem}

\usepackage{array,amsmath,amsthm,amssymb,amsfonts,graphicx,setspace,color,verbatim,wrapfig,hyperref,mathtools,stackengine}

\title{}
\date{} 

\newcommand{\Z}{\mathbb{Z}}

\newcommand{\nn}[2]{\langle #1 #2 \rangle}

\renewcommand{\d}{\partial}

\renewcommand{\nn}{\nonumber}

\newcommand{\beq}{\begin{equation}}
\newcommand{\eeq}{\end{equation}}
\newcommand{\ba}{\begin{array}{ccc}}
\newcommand{\ea}{\end{array}}
\newcommand{\bea}{\begin{eqnarray}}
\newcommand{\eea}{\end{eqnarray}}

\newcommand{\TNK}{{\cal T}_{NK}}
\newcommand{\T}{{\cal T}}

\graphicspath{{figures/}}

\begin{document}

\title{A 1d lattice model for the boundary of the quantum spin-Hall insulator}

\author{Max A. Metlitski}
\affiliation{Department of Physics, Massachusetts Institute of Technology, Cambridge, MA  02139, USA}

\date{\today}

\begin{abstract}
We present a 1d lattice model that mimics the boundary of the conventional 2d quantum spin-Hall  insulator (QSHI) with $U(1)$ symmetry and time-reversal ${\cal T}$, satisfying ${\cal T}^2 = (-1)^F$. Our construction utilizes a  local tensor product Hilbert space of finite site dimension with a non-onsite symmetry action. We discuss how several signature properties of the QSHI, such as the fractional charge on ${\cal T}$-domain walls and Kramers parity  switching upon $\pi$-flux threading, are manifested in our treatment. We also present a 1d Hamiltonian whose ground state realizes the conventional Luttinger-liquid phase of the QSHI edge.

\noindent

\end{abstract}

\maketitle

\tableofcontents

\newpage

\section{Introduction}

Symmetry protected topological (SPT) phases have attracted a lot of attention in recent years.\cite{Chen2012, Chen2013, SenthilSPTRev}  The most striking feature of these phases is the existence of non-trivial ``anomalous" edge states. By anomalous, it is meant that the edge cannot be re-created without the bulk under the standard physical assumptions: a local tensor product Hilbert space $V = \otimes_i V_i$, and an onsite action of the symmetry $G$, $U(g) = \prod_i U_i(g)$, where $i$ labels the boundary sites and $U_i(g)$ is a unitary acting on site $i$ satisfying the group law $U_i(g) U_i(h) = U_i(gh)$. However, it was realized early on that for a class of SPT phases, namely,  SPT phases in the group cohomology classification, the edge can, in fact, be mimicked without the bulk, provided that one relaxes the assumption of onsite symmetry action.\cite{Chen2012, CZX, CX, Levin_Gu} Instead, one takes $U(g)$ to act as a finite depth local unitary that satisfies the group law only globally. Furthermore, at least in 1d,\footnote{Here and below lower $d$ always stands for spatial dimension.} given $U(g)$ one can extract algebraic data characterizing the ``non-onsiteness" that precisely matches the  data labeling the bulk 2d SPT.\cite{ElseNayak} For instance, for 2d SPTs of bosons this data is a cocycle $w_3 \in H^3(G, U(1))$. 

It should be noted that not all SPT phases are covered by the group cohomology classification (or its fermionic ``supercohomology" generalization).\cite{Vishwanath2013, Burnell2013, KapustinBos, supercohomology, GK, Fidkowski2014} In fact, it is understood that the boundaries of some beyond (super)cohomology phases cannot be mimicked by giving up just the onsite symmetry action.\cite{ElseNayak, MJ} Instead, in some examples the boundary can be mimicked by further relaxing the assumption of a local tensor product Hilbert space and  working in a Hilbert space that is constrained.\cite{SonHLL,ChongDual, MVDual, SenthilHLL, flatband, MJ}

Returning to phases in group cohomology, a subtlety exists when the group $G$ is not finite. Indeed, the canonical construction of the exactly solvable bulk Hamiltonian and the effective boundary model with a non-onsite symmetry utilizes a site Hilbert space labeled by  group elements $g$.\cite{Chen2013} Thus, if $G$ is continuous, the site Hilbert space dimension is infinite. It is not obvious how to truncate this Hilbert space to a finite dimensional one, especially since the cochain entering the construction of the non-onsite symmetry is generally a discontinuous function of the group variables in this case. Thus, it remains an open question whether for continuous group $G$ for phases in group (super)cohomology the boundary can be mimicked with a  Hilbert space of finite site dimension.

In this paper, we answer the above question for the case of the conventional 2d quantum spin Hall insulator (QSHI) protected by the $U(1)$ particle number symmetry and time-reversal ${\cal T}$ with ${\cal T}^2 = (-1)^F$.\cite{KaneMele2d} According to Ref.~\onlinecite{WenSuper2018}, this phase is within the (generalized) supercohomology classification, however, the continuos $U(1)$ symmetry leads to the difficulties mentioned above. We show that in this case the 1d boundary can, indeed, be mimicked with a local tensor product fermionic Hilbert space of finite site dimension at a cost of a non-onsite symmetry action.  We will demonstrate how the non-onsite symmetry action leads to several key properties of the QSHI edge such as:
\begin{enumerate} 
\item fractional electric charge $n+1/2$ on ${\cal T}$ domain walls (here $n$ is an integer); 
\item switching of Kramers parity from ${\cal T}^2 = (-1)^{F}$ to ${\cal T}^2 = - (-1)^F$ upon threading flux $\pi$ through the ring. 
\end{enumerate}
We will also extract the algebraic data characterizing our non-onsite symmetry. Finally, we will present a 1d lattice Hamiltonian that realizes the conventional Luttinger-liquid phase of the QSHI edge.

In fact, it proves convenient to initially work with a slightly larger symmetry group than that of a conventional QSHI. We will consider a 2d fermion SPT with the following set of symmetries: 
\begin{enumerate}
\item $U(1)$ particle number with a corresponding charge $N$, so that fermion parity $(-1)^F = (-1)^N$;
\item A unitary $\mathbb{Z}_2$ symmetry $U$;
\item An anti-unitary $\mathbb{Z}_2$ symmetry ${\cal T}_{NK}$,
\end{enumerate}
satisfying the following algebra:
\beq U^2 = 1, \quad {\cal T}^2_{NK} = 1, \quad [U, N] = 0, \quad [\TNK, N] = 0, \quad {\cal T}_{NK} U = (-1)^N U {\cal T}_{NK}.  \label{symalg}\eeq
The subscript on ${\cal T}_{NK}$ stands for non-Kramers. We then define the Kramers time-reversal symmetry as 
\beq \T = U \TNK.\eeq
Then
\beq \T^2 = (-1)^F, \quad [\T, N] = 0.\eeq
The symmetry group $(U(1)\rtimes \mathbb{Z}^T_4)/\mathbb{Z}_2$ of QSHI is obtained by keeping just the $U(1)$ symmetry and the Kramers time-reversal $\T$. It is, nevertheless, convenient for our discussion to initially consider the larger symmetry group introduced above - when needed, it can be broken down to the physical subgroup. 

The SPT we consider can be visualized as follows. We stack a layer of integer quantum Hall effect with $\sigma_{xy} = 1$ and a layer of integer quantum Hall effect with $\sigma_{xy} = -1$. The fermions in the $\sigma_{xy} = 1$ layer are taken to be charged under the unitary $\mathbb{Z}_2$ $U$, while the fermions in the $\sigma_{xy} = -1$ layer are neutral under $U$. Fermions in both layers carry $U(1)$ charge of $1$. $\TNK$ simply interchanges the fermions in the two layers: $c^+ \leftrightarrow c^-$.  

The edge is then a  $1+1$D Dirac fermion,
\beq H_{edge} = \int dx  \left(\psi^{\dagger}_R (-i \d_x) \psi_R + \psi^{\dagger}_L (i \d_x) \psi_L\right), \label{psiedge}\eeq
with 
\beq U: \quad \psi_R \to \psi_R, \quad \psi_L \to -\psi_L, \label{Z2psi}\eeq
and both $\psi_R, \psi_L$ carry charge $1$ under $U(1)$. Further,
\beq \TNK: \quad \psi_R \leftrightarrow \psi_L, \quad i \to -i.\eeq
We see that the combination $\T$ acts on the edge exactly like time-reversal in the physical QSHI does. 

In the first part of this paper, section \ref{sec:unconstr}, we mimic the 1d edge of the SPT above using a non-onsite symmetry implementation. Furthermore, applying a Jordan-Wigner (JW) transformation, we   re-write the edge as a 1d bosonic $\mathbb{Z}_2$ gauge theory. In this formulation, it becomes simple to write down and analyze symmetric edge Hamiltonians. We note that this bosonized formulation is very similar to that applied in the pioneering work, Ref.~\onlinecite{supercohomology}, to the edge of the SPT with just unitary $\mathbb{Z}_2$ symmetry $U$.

In the second part of the paper, section \ref{sec:constr},  we  present an exactly solvable bulk 2d Hamiltonian for the SPT with just $U(1)$ and unitary $\mathbb{Z}_2$ symmetry $U$. We  study the edge of this Hamiltonian and introduce a ``bosonized" labeling for the edge Hilbert space. This bosonized labeling and the action of $U(1)$ and $\mathbb{Z}_2$ symmetry within it matches the JW transformed theory from the first part of the paper subject to a further local constraint that can be enforced energetically. Thus, our somewhat ad-hoc construction in section \ref{sec:unconstr} matches the bulk+boundary construction of section \ref{sec:constr}. 

While this paper was being completed, Ref.~\onlinecite{Alicea} appeared that, among other results, also discusses how to mimic the boundary of a 2d QSHI in a 1d lattice model. One difference with the treatment in the first part of our paper is that the symmetry action of Ref.~\onlinecite{Alicea} only satisfies the group algebra in a constrained Hilbert space - this is somewhat akin to the 1d model for the edge of beyond supercohomology SPTs in Ref.~\onlinecite{MJ}.  On the other hand, our non-onsite symmetry action in section \ref{sec:unconstr} satisfies the group algebra in the entire Hilbert space without the need for constraints. We also note that Ref.~\onlinecite{Alicea} presents an exactly solvable model for the bulk of the QSHI and derives the 1d edge model starting from this bulk. This is similar to section \ref{sec:constr} of our paper; however, we only give such a bulk+boundary construction for the simpler case of $U(1)$ and unitary $\mathbb{Z}_2$ symmetry and not time-reversal. 

We would also like to direct the reader's attention to appendix \ref{app:Ugen}, which has a somewhat different focus from the rest of the paper. There we present a non-onsite boundary symmetry action  for any 2d supercohomology fermion SPT with a finite symmetry group $G_f$.

\section{1d model}
\label{sec:unconstr}

\subsection{Symmetry action}

\label{sec:symmact}
We now describe a 1d model that mimics the edge (\ref{psiedge}). We take a chain of $L$ sites arranged on a ring with fermions living on each site. The  fermion creation, annihilation operators $c_i$,$c^{\dagger}_i$, $i = 1 \ldots L$, satisfy the standard algebra $\{c_i, c^{\dagger}_j\} = \delta_{ij}$. It will be convenient to define the Majorana operators $\gamma_i$,$\bar{\gamma}_i$:
\beq c_i = \frac12(\gamma_i + i \bar{\gamma}_i), \quad c^{\dagger}_i = \frac12(\gamma_i - i \bar{\gamma}_i). \label{cgamma}\eeq
We further introduce Ising spin variables living on each link $(i,i+1)$ of the chain. We will denote the corresponding Pauli matrices by $\tau^a_{i,i+1}$, with $a = 1,2,3$. We will commonly work in the $\tau^z$ basis and sometimes use the notation $\tau^z_{i,i+1} = (-1)^{g_{i,i+1}}$ with $g_{i,i+1} \in \{0,1\}$.

We define the symmetry operations as follows. First, the fermion number is defined as
\beq N = \sum_j n_j, \quad n_j = (-1)^{g_{j-1,j} (g_{j,j+1} + 1)}\, c^{\dagger}_j c_j. \label{Ndef} \eeq
That is $n_j = - c^{\dagger}_j c_j$ if $j$ is at the domain wall between a ``$-$'' Ising spin on the left and a ``$+$" Ising spin on the right. Otherwise, $n_j = c^{\dagger}_j c_j$. Note that $n_j$ has integer eigenvalues, further, $(-1)^{n_j} = (-1)^{c^{\dagger}_j c_j}$. Thus, $(-1)^N$ is the standard fermion parity operator. 

Next, we define the generator of the unitary $\mathbb{Z}_2$ symmetry,
\beq U = \left(\prod_{j=1}^{L} \tau^x_{j,j+1}\right) \left(\prod_{j = 1}^L \gamma_j^{g_{j-1,j} + g_{j,j+1}}\right) (-i)^{N_{dw}/2}. \label{Udef}\eeq
This form was inspired by Eq.~(19) of Ref.~\onlinecite{NatAshvin} and also the general discussion in appendix \ref{app:Ugen}. Here $N_{dw}$ is the number of domain walls in the $\tau^z$ configuration,
\beq N_{dw} = \sum_{j=1}^L \frac{1-\tau^z_{j-1,j} \tau^z_{j,j+1}}{2}.\eeq
Note that $N_{dw}$ is always an even integer. We also note that the terms in the $\gamma$ product in $(\ref{Udef})$ generally do not commute. We use a definition where terms with smaller $j$ appear to the left. Note that $U$ is a fermion parity even operator. Moreover, it is locality preserving:
\bea U \gamma_j U^{\dagger} &=& (-1)^{g_{j-1,j} + g_{j,j+1}} \gamma_j, \nn\\
U \bar{\gamma}_j U^{\dagger} &=& \bar{\gamma}_j, \nn\\
U \tau^z_{j,j+1} U^{\dagger} &=& -\tau^z_{j,j+1},\nn\\
U \tau^x_{j,j+1} U^{\dagger} &=& \tau^x_{j,j+1} (-i s_{j,j+1} \gamma_j \gamma_{j+1}) (-1)^{g_{j,j+1}(g_{j-1,j} + g_{j+1,j+2})} i^{g_{j+1,j+2}-g_{j-1,j}}. \label{Uopac} \eea
Here $s_{L,1} = -1$ and all other $s_{j,j+1} = 1$. We can think of $s$ as a spin-structure; the above $s$ corresponds to Neveu-Schwarz (NS) spin structure (anti-periodic boundary conditions for the fermions) with a ``branch-cut" across the link $(L,1)$. One way to see this is from the commutation of $U$ with the translation operator. Define $T_x$ to be  the translation by one to the right,  so that $T_x c_i T^{\dagger}_x = c_{i+1}$ and $T_x \tau^a_{i,i+1} T^{\dagger}_x = \tau^a_{i+1,i+2}$. Then 
\beq [(-1)^{n_1} T_x, U]  = 0.\eeq
Thus, it is $(-1)^{n_1} T_x$ which commutes with $U$ - a translation followed by a gauge transformation necessary to move the branch-cut back into place.

Finally, we define the anti-unitary operator $\TNK$ as follows. First, let ${\cal T}_0$ be the anti-unitary operator that sends
\bea && {\cal T}_0 c_j {\cal T}^{\dagger}_0 = c_j, \quad i \to -i, \nn \\
&& {\cal T}_0 \tau^{x,z}_{j,j+1} {\cal T}_0 = \tau^{x,z}_{j,j+1}, \quad  {\cal T}_0 \tau^{y}_{j,j+1} {\cal T}_0 = - \tau^{y}_{j,j+1}, \label{T0} \eea
i.e. as far as the Ising spin variables are concerned, ${\cal T}_0$ simply acts by complex conjugation in the $\tau^z$ basis. Now define,

\beq \TNK = \left(\prod_{j=1}^L (-1)^{g_{j,j+1} \, c^{\dagger}_j c_j} \right) {\cal T}_0. \label{TNKdef} \eeq
Again, $\TNK$ is locality preserving:
\beq \TNK c_i \TNK^{\dagger} = (-1)^{g_{i,i+1}} c_i, \quad \TNK \tau^z_{i,i+1} \TNK^{\dagger} = \tau^z_{i,i+1}, \quad \TNK \tau^x_{i,i+1} \TNK^{\dagger} = (-1)^{c^{\dagger}_i c_i} \tau^x_{i,i+1}.\eeq

One can check that $N$, $U$ and $\TNK$, indeed, satisfy the algebra (\ref{symalg}).

\subsection{Domain Walls}
We now discuss how our 1d lattice model reproduces the structure of domain walls in the continuum edge theory (\ref{psiedge}).
Imagine one turns on a mass 
\beq \delta H_m = m (\psi^{\dagger}_R \psi_L +  \psi^{\dagger}_L \psi_R), \eeq
gapping out the edge modes. Under $U$ and $\T$, $m \to -m$. Now consider a domain wall between regions with $m > 0$ and $m < 0$. Solving for the spectrum we find a Dirac zero mode localized at the domain wall. When the zero mode is filled (empty), the domain wall has a  charge $1/2$ ($-1/2$) localized in its vicinity. Thus, while the symmetries we are considering do not pin the chemical potential to zero, the domain wall carries charge $N \in {\mathbb{Z}} + 1/2$. How does this manifest itself in terms of the global charge of the system (say, on a ring)? We can let the segment $x \in (a, b)$ have $m < 0$ and the complement of the segment have $m > 0$. If the domain walls at $x = a$ and $x = b$ are symmetry conjugates (under $U$ or $\T$) of each other, then they will carry identical charge. Thus, the total charge of the system will be an odd integer. 

Let's see how this effect plays out in our 1d lattice model. Consider a Hamiltonian
\beq H = -\sum_i m_{i,i+1} \tau^z_{i,i+1} + \sum_i \frac{1+\tau^z_{i-1,i} \tau^z_{i,i+1}}{2} c^{\dagger}_i c_i.  \label{mlat}\eeq  
Here $m_{i,i+1}$ is a real number. Under $U$ and $\T$, $m_{i,i+1} \to -m_{i,i+1}$. The first term in (\ref{mlat}) pins the Ising spin to the local sign of $m$. The second term pins the fermion occupation number $c^{\dagger}_i c_i$ to zero if there is no domain wall at $i$. Thus, if $m$ is uniform, the ground state is unique and gapped.   Now, consider a configuration of $m$ where in the vicinity of $i = a$,  $m_{i,i+1} > 0$ for $i < a$ and $m_{i, i+1} < 0$ for $i \ge a$. There is now a two-fold degeneracy in the spectrum of (\ref{mlat})  associated with the occupation number $c^{\dagger}_a c_a = 0,1$.  The symmetries we are considering do not protect this degeneracy. However, under $U$ and $\T$, the sign of $m$ flips and also the occupation number $c^{\dagger}_a c_a$ flips due to (\ref{Uopac}). Now, imagine we have two domain walls: one at $i  =a$ and one at $i  =b$: $m_{i,i+1} < 0$ for $a \leq i \leq b-1$ and $m_{i,i+1} > 0$ otherwise. If the domain walls at $i = a$ and $i = b$ are related by $U$ or $\T$ then if $c^{\dagger}_a c_a = 1$ then $c^{\dagger}_b c_b  =0$, while if $c^{\dagger}_a c_a = 0$ then $c^{\dagger}_b c_b = 1$. In the first case, we find the global charge of the system, (\ref{Ndef}), $N  = 1$, while in the second case $N  = -1$. Thus, in both cases the global charge of two symmetry-related domain walls is an odd integer, as expected. 

\subsection{Flux-threading and anomalies}
We now discuss a thought experiment that demonstrates the anomaly of the edge theory (\ref{psiedge}). Imagine threading flux $\phi$ of $U(1)$ symmetry through the ring. Upon threading flux $2\pi$ the $\mathbb{Z}_2$ charge in the theory (\ref{psiedge}) flips. A related point is that while with NS boundary conditions we have ${\cal T}^2 = (-1)^F$, once we thread flux $\phi = \pi$ through the ring, ${\cal T}^2 = - (-1)^F$. How are these anomalies manifested in our 1d lattice formulation?

First, we have to discuss how to thread flux of $U(1)$ symmetry. We note that the charge $N$, (\ref{Ndef}), is a sum of local commuting operators $n_j$ with integer eigenvalues. Thus, while the $U(1)$ transformation $e^{i \alpha N}$ is not a strictly onsite symmetry in our formulation, the model can be coupled to a background gauge field as follows. Say we want to thread flux $\phi$ through the link $(L,1)$. Then, for any operator $O$ in the Hamiltonian localized near this link, we replace $O$ by $S(\phi) O S^{\dagger}(\phi)$ where 
\beq S(\phi) = e^{i \phi \sum_{j = 1}^{p} n_j}, \label{Sphi}\eeq
with $p$ - a number much larger than the support of $O$. Operators localized far from the $(L,1)$ link are not conjugated. However, this is not entirely satisfactory since this procedure breaks the $\mathbb{Z}_2$ symmetry $U$. Indeed, while $[N, U] = 0$, $n_i$ itself is not $\mathbb{Z}_2$ invariant. Indeed, 
\beq U n_i U^{\dagger} = n_i + \frac12(\tau^z_{i,i+1}-\tau^z_{i-1,i}).\eeq
However, we may instead define
\beq \tilde{n}_i = n_i + \frac{1}{4}(\tau^z_{i,i+1} - \tau^z_{i-1,i}). \eeq
Then the total charge $N = \sum_i \tilde{n}_i$, but now $[U, \tilde{n}_i] = 0$, so the Hamiltonian commutes with $U$ for any flux $\phi$. The cost one pays for this is that $\tilde{n}_i$ now  has half-integer eigenvalues. As a result, if we implement flux insertion with $\tilde{S}(\phi)$ defined analagosly to Eq.~(\ref{Sphi}), but with $n_j$ replaced by $\tilde{n}_j$ in the exponent,  $\tilde{S}(\phi + 2\pi) \neq \tilde{S}(\phi)$. Instead, $\tilde{S}(\phi + 2\pi) = \tau^z_{L,1} \tau^z_{p,p+1} \tilde{S}(\phi)$. Since we only conjugate terms in the Hamiltonian localized near the branch cut, 
\beq H_{\phi+2\pi} = \tau^z_{L,1} H_{\phi} \tau^z_{L,1}. \label{phiper}\eeq
Now, imagine we start threading flux though the ring. Let the corresponding time-evolved wave-function be $|\psi(\phi)\rangle$.  Since $H_\phi$ commutes with $U$, the $U$ charge of $|\psi(\phi)\rangle$ does not depend on $\phi$. However, if we want to meaningfully compare the wave-functions at flux $\phi$ and $\phi + 2\pi$, due to $(\ref{phiper})$, we have to compare $|\psi(\phi)\rangle$ and $\tau^z_{L,1} |\psi(\phi+2\pi\rangle$. Since $\tau^z_{L,1}$ anticommutes with $U$, these will have opposite $U$ charge. This demonstrates the implementation of anomaly in our lattice treatment. 

We can, similarly, discuss the Kramers parity switching between $\phi = 0$ and $\phi = \pi$. We note that $\TNK \tilde{n}_i \TNK^{\dagger} = \tilde{n}_i$. Therefore, $\TNK H_\phi \TNK = H_{-\phi}$ and likewise, $\T H_\phi \T = H_{-\phi}$, where we recall $\T = U \T_{NK}$.  Thus, generally, flux insertion breaks time-reversal symmetry. However, when $\phi = \pi$, $\TNK H_{\pi} \TNK = H_{-\pi} = \tau^z_{L,1} H_{\pi} \tau^z_{L,1}$. Thus, $[\tau^z_{L,1} \TNK, H_{\pi}] = 0$. Likewise, $\T' = U \tau^z_{L,1} \TNK$ is a symmetry when $\phi = \pi$. But, $(\T')^2 = (U \tau^z_{L,1} \TNK)^2 = - (U \TNK)^2 = - (-1)^F$. This is, indeed, the expected result. 

\subsection{Jordan-Wigner transformation}
\label{sec:JW}
We now apply the Jordan-Wigner (JW) transformation to our 1d fermion model. This will allow us to construct and analyze symmetric Hamiltonians that at low-energy realize the theory (\ref{psiedge}). We begin with the standard JW transformation:
\beq c_i = \exp\left(\pi i \sum_{j=1}^{i-1}(1+ \mu^z_j)/2\right) \mu^-_i, \quad\quad c^{\dagger}_i c_i = \frac{1+\mu^z_i}{2}, \label{JW} \eeq
where $\mu^a_i$ are Pauli operators living on each site. The symmetry operators become:
\bea 
N &=& \sum_i \tilde{n}_i, \quad\quad \tilde{n}_i = \frac{1}{4}(1+\tau^z_{i-1,i} \tau^z_{i,i+1})(1+\mu^z_i) - \frac14 (\tau^z_{i,i+1} -\tau^z_{i-1,i}) \mu^z_i, \nn\\
U &=& \left(\prod_{j =1}^{L} \tau^x_{j,j+1}\right) (-i)^{N_{dw}/2} \left(\prod_{j=1}^{L} (\mu^x_j)^{g_{j-1,j} + g_{j,j+1}}\right) \left(\prod_{j = 1}^L (-\mu^z_j)^{g_{j,j+1}} \right) (-1)^{N g_{L,1}}, \nn\\
\TNK &=& \prod_j (-\mu^z_j)^{g_{j,j+1}} K. \eea
where the complex conjugation $K$ in $\TNK$ is performed in the $\tau^z$, $\mu^z$ basis. One can simplify the above forms with a unitary transformation:
\beq R =  \left(\prod_j e^{\frac{\pi i}{4} (1+ \mu^z_j) g_{j,j+1}}\right) \left(\prod_j ({\mu^x_j})^{g_{j-1,j} (g_{j,j+1} +1)}\right) (-i)^{[N] g_{L,1}}. \label{Rdef}\eeq
Here and below $[x] = 0$ if $x$ is even and $[x] = 1$ if $x$ is odd. For an operator $O$, letting $\hat{O} = R O R^{\dagger}$, we have
\bea \hat{N} &=& \sum_i \hat{\tilde{n}}_i, \quad\quad \hat{\tilde{n}}_i = \frac{1+\mu^z_i}{2} - \frac14(1-\tau^z_{i-1,i} \tau^z_{i,i+1}), \nn\\
\hat{U} &=& i^{\hat{N} - [\hat{N}]} \prod_j \tau^x_{j,j+1}, \nn\\
\hat{\T}_{NK} &=& (-1)^{g_{L,1} \hat{N}} K. \label{SymmJW}\eea
We see that the symmetry effectively acts slightly differently in the even and odd fermion parity sectors. 

Let's now discuss local boson operators of the original fermion theory in this ``bosonized" treatment. (We discuss local fermion operators in  appendix \ref{app:ferm}.) 
Recall, there is  a subtlety with the JW transformation on the circle: while, in general, local boson operators in the fermion theory map to local boson operators with even $N$ in the bosonized theory, this is not strictly true for operators localized near the $(L,1)$ link. Rather, for these operators the bosonized form reduces to two generally different sets of local operators, depending on whether the state they are acting on has even or odd $N$: $O_{odd} = Q O_{even} Q^{\dagger}$, where $Q$ is a $\pi$-flux threading operator
\beq Q = \prod_{j = 1}^{p} (-\mu^z_j),\eeq
with $p$ - a number much greater than the support of the local operator $O$. The same is true after we perform the unitary rotation  (\ref{Rdef}): $\hat{O}_{odd} = \hat{Q} \hat{O}_{even} \hat{Q}^{\dagger}$ with 
\beq \hat{Q} = \prod_{j = 1}^{p} e^{-\pi i \hat{\tilde{n}}_j}. \label{hatQ} \eeq

It will be convenient to formulate the above bosonized theory as a $\mathbb{Z}_2$ gauge theory. Consider the operator $\hat{\tilde{n}}_i$. It has eigenvalues $0$, $1$ and $\pm \frac12$. The eigenvalues $0$ and $1$ are realized if there is no domain wall at $i$, while the eigenvalues $\pm \frac12$ are realized if there is a domain wall at $i$. Thus, we may think of $\hat{\tilde{n}}_i$ as independent variables, provided we impose the constraint $G_i \sim 1$ with
\beq G_i = (-1)^{2 \hat{\tilde{n}}_i} \tau^z_{i-1,i} \tau^z_{i,i+1}. \label{eq:Gauss} \eeq
We may then think of the system as a boson coupled to a $\mathbb{Z}_2$ gauge field. The boson density is given by $2 \hat{\tilde{n}}_i$ and is allowed to take on values $-1$, $0$, $1$ and $2$, and the total boson number is $N_b = \sum_i  2 \hat{\tilde{n}}_i$.  Eq.~(\ref{eq:Gauss}) becomes the Gauss law constraint, with $\tau^z_{i,i+1}$ playing the role of $\mathbb{Z}_2$ electric field on link $(i,i+1)$. Then $\tau^x_{i,i+1}$ plays the role of $\mathbb{Z}_2$ vector potential. The quantity $\prod_i \tau^x_{i,i+1}$ is  the $\mathbb{Z}_2$ gauge flux around the ring - a gauge invariant quantity. Note that this quantity enters the $\mathbb{Z}_2$ symmetry $\hat{U}$, Eq.~(\ref{SymmJW}).  The physical fermion number $\hat{N}$ is half the boson number $N_b$.  The Gauss law (\ref{eq:Gauss}) guarantees that the total boson number $N_b$ is even, so $\hat{N}$ is an integer.

We now give an example of a Hamiltonian that commutes with the symmetries (\ref{SymmJW}). It will be convenient to dynamically suppress the state $2 \hat{\tilde{n}}_i = 2$, i.e. enforce that in the absence of a domain wall at $i$, $\hat{\tilde{n}}_i = 0$ and so $c^{\dagger}_i c_i = 0$. This can be done with a term
\beq H_{u} = u \sum_i \hat{\tilde{n}}_i (1+ \tau^z_{i-1,i} \tau^z_{i,i+1}), \label{HU} \eeq
with $u > 0$. Then in the ground state subspace of $H_u$, $2 \hat{\tilde{n}}_i$ takes values $-1, 0, 1$ and we can think of our system as a spin $1$ chain. We let $S^a_i$ be spin 1 operators localized on each site,  $S^+ = \sqrt{2} \left(\begin{array}{ccc} 0 & 1 &0 \\ 0 & 0 & 1\\ 0 & 0 & 0\end{array}\right),\,\, S^-  = \sqrt{2} \left(\begin{array}{ccc} 0 & 0 &0 \\ 1 & 0 & 0\\ 0 & 1 & 0\end{array}\right), \,\, S^z  = \left(\begin{array}{ccc} 1 & 0 &0 \\ 0 & 0 & 0\\ 0 & 0 & -1\end{array}\right)$,
with $S^z_i = 2 \hat{\tilde{n}}_i$. The Gauss constraint (\ref{eq:Gauss}) in the $H_u$ ground state subspace becomes
\beq G_i = (-1)^{S^z_i} \tau^z_{i-1,i} \tau^z_{i,i+1}, \label{Gi}\eeq
and the symmetries are given by,
\beq N = \frac{S^z}{2},\eeq
\bea  U &=& \exp\left(\frac{\pi i}{4} S^z\right) \prod_i \tau^x_{i,i+1} \times \left\{ \begin{array}{ll} 1, & N \,\,-\,\, even\\ -i, & N \,\,-\,\, odd,\end{array} \right. \label{US}\eea
\bea  {\cal T}_{NK} = \left\{ \begin{array}{ll} K, &\,\,\,\, N \,\,-\,\, even\\ \tau^z_{L,1} K, & \,\,\,\, N \,\,-\,\, odd.\end{array} \right. \label{TNKS}\eea
Here $S^z = N_b = \sum_i S^z_i$ is the total boson number;  we drop hats on $N$, $U$ and $\TNK$ here and below.

We may consider the Hamiltonian $H  = H_u + H_{S  =1}$ with $H_{S = 1} = \sum_i H_{i,i+1}$ and
\beq  H_{i,i+1} = -\frac{J}{2} (S^+_i S^-_{i+1} + S^-_i S^+_{i+1}) \tau^{x}_{i,i+1}, \quad N - {\rm even}. \label{eq:Hspin1}\eeq
For odd $N$, $H_{i,i+1}$ is still given by Eq.~(\ref{eq:Hspin1}) for $i \neq L$, while for $i  = L$,
\beq H_{L,1} = - \frac{J}{2} (i S^+_{L} S^-_{1} - i S^-_{L} S^+_{1}) \tau^{x}_{L, 1}, \quad N  - {\rm odd}. \label{eq:Hspin2}\eeq
This Hamiltonian commutes with the Gausses law constraint (\ref{Gi}) and with all symmetries. Note that for even and odd $N$ the terms in the Hamiltonian near the branch cut are, indeed, related by the flux-threading operator (\ref{hatQ}), $\hat{Q} = \prod_{j = 1}^{p} e^{-\pi i S^z_j/2}$. From the boson point of view this threads flux $\pi/2$ through the cut. For completeness, we give the form of the above Hamiltonian in the original fermionic variables in appendix \ref{app:fermH}. 

We observe that the Hamiltonian (\ref{eq:Hspin1}) is just a spin 1 XX chain coupled to a $\mathbb{Z}_2$ gauge field. We may also add a ``ZZ" term preserving the symmetry:
\beq \delta H_{i,i+1} = -\Delta S^z_i S^z_{i+1}. \label{DeltaS1} \eeq
In the absence of the $\mathbb{Z}_2$ gauge field, the Hamiltonian above is numerically known to be in the Luttinger liquid phase for $\Delta_c < \Delta < |J|$, with $\Delta_c/|J|$ lying close to $0$.\cite{JullienS1,MinoruS1, KiyomiS1} We explain in section \ref{cont} how turning the $\mathbb{Z}_2$ gauge field on and treating correctly the boundary conditions in the odd $N$ sector (\ref{eq:Hspin2}) gives the standard Luttinger liquid theory for the QSHI edge (\ref{psiedge}) with the correct action of symmetries $U$ and $\TNK$. 

 If we like, we can place a further energetic constraint to reduce the Hamiltonian to a spin 1/2  chain (hardcore boson) coupled to a $\mathbb{Z}_2$ gauge field. Indeed, we may add a term that penalizes the $S^z_i = -1$ state, e.g. $\delta H = u \sum_i S^z_i (S^z_i - 1)$. Then for large $u$, $S^z_i$ effectively only takes two values $S^z_i = 0$ and $S^z_i = 1$. Thinking in the hardcore boson language, $S^z_i = a^{\dagger}_i a_i$, $[a_i, a^{\dagger}_j] = \delta_{ij}$, the Hamiltonian (\ref{eq:Hspin1}) projected onto $S^z_i \neq -1$ subspace becomes
\bea H_{i,i+1} &=& - J (a^{\dagger}_i a_{i+1} + a^{\dagger}_{i+1} a_i) \tau^x_{i,i+1},  \quad N - {\rm even} \label{hardcore},\\
 H_{L,1} &=& - J (i a^{\dagger}_{L} a_{1} - ia_{L} a^{\dagger}_{1}) \tau^{x}_{L, 1}, \quad\,\,\,\,\,\, N  - {\rm odd} \label{hardcore2}.\eea
Symmetry also allows us to add a ``ZZ" coupling
\beq \delta H_{i,i+1} = -  2 \Delta (a^{\dagger}_i a_i - \frac12) (a^{\dagger}_{i+1} a_{i+1} -\frac12). \label{ZZ}\eeq
The Hamiltonian above is  exactly solvable by Bethe ansatz and, in the absence of a $\mathbb{Z}_2$ gauge field, is known to be in the Luttinger liquid phase for $|\Delta| < |J|$. Again, turning on the $\mathbb{Z}_2$ gauge field gives the correct QSHI edge theory (\ref{psiedge}). 

We conclude this section by noting that the reason we chose to discuss the spin 1 Hilbert space  without going immediately to the more ``economical" spin 1/2 Hilbert space is that the former matches the effective lattice edge model we derive starting from the bulk construction in section \ref{sec:constr}. 

\subsection{Luttinger liquid description}
\label{cont}
We now argue that the lattice construction above, indeed, correctly describes the QSHI edge. For definiteness, we work with the $S = 1$ Hamiltonian (\ref{eq:Hspin1}), (\ref{eq:Hspin2}), (\ref{DeltaS1}) although our conclusions are much more general.\footnote{In particular, the analysis and conclusions apply to the exactly solvable $S=1/2$ Hamiltonian (\ref{hardcore}), (\ref{hardcore2}), (\ref{ZZ}), with minimal modifications related to the finite  boson density in the ground state.} First, consider the same Hamiltonian, but with the $\mathbb{Z}_2$ gauge field turned off:
\beq H = -\sum_i \left(\frac{1}{2} (S^+_i S^-_{i+1} + S^-_i S^+_{i+1}) + \Delta S^z_i S^z_{i+1}\right). \eeq
Here, we've set $J = 1$. This is a Hamiltonian of bosons with a $U(1)$ symmetry. One possible phase it can be in is a Luttinger liquid. In fact, this is numerically known to be the case for $\Delta_{c} < \Delta < 1$ with $\Delta_c$ close to 0.\cite{JullienS1,MinoruS1, KiyomiS1}  An effective continuum theory describing this Luttinger liquid is
\beq L = \frac{1}{2\pi} \d_x \tilde{\phi} \d_t \tilde{\theta} - \frac{1}{2\pi} \left(\tilde{K} (\d_x \tilde{\phi})^2 + \frac{1}{4\tilde{K}} (\d_x \tilde{\theta})^2\right).\eeq
Here and below we set the speed of the excitations to $1$. Here, 
\beq S^+_j \sim e^{i \tilde{\phi}}, \quad S^z_j = \frac{1}{2\pi} \d_x \tilde{\theta} + \ldots \quad.\eeq
The total spin $S^z = \frac{1}{2\pi}(\tilde{\theta}(x+L) -\tilde{\theta}(x))$. $\tilde{K}$ is the Luttinger parameter. The reason for the tildes will become clear shortly. 
Under time-reversal symmetry acting as complex-conjugation in the $S^z$ basis,
\beq  K: \,\,\tilde{\phi} \to -\tilde{\phi}, \quad\quad \tilde{\theta} \to \tilde{\theta}.\eeq

Next, we turn the $\mathbb{Z}_2$ gauge field back on. We have two flux sectors, $e^{i \Phi} = \prod_i \tau^x_{i,i+1} = \pm 1$, i.e. $\Phi = 0, \pi$. We note that all $\tau^x_{i,i+1}$ commute with the Hamiltonian (\ref{eq:Hspin1}), (\ref{eq:Hspin2}), (\ref{DeltaS1}). 
We may, thus, fix a gauge for all $\tau^x_{i, i+1}$, find the eigenstates $|\psi\rangle$ of $H$,  and then obtain an eigenstate satisfying the Gausses law: $\prod_i(1+G_i) |\psi\rangle$, with $G_i$ given by (\ref{Gi}). 
This gives a non-vanishing state provided that total $S^z$ is even. Thus, when $e^{i \Phi} = 1$, we may set all $\tau^x_{i,i+1} = 1$, and when $e^{i \Phi} = -1$, we may set say $\tau^x_{L,1} = -1$ and all the other  $\tau^x$'s to $+1$. 

We must also remember that the Hamiltonian in the even and odd $N$ sectors is different. Thus, in the even $N$ sector, $S^z = 0 \quad (mod\,\, 4)$,  the boson $e^{i \tilde{\phi}}$ sees a flux $e^{i \Phi} = \pm 1$, while in the odd $N$ sector, $S^z = 2 \quad (mod\,\, 4)$, the boson sees a flux $-i e^{i \Phi} = \mp i$.
This translates into the continuum theory as: 
\bea \tilde{W} &\in&  {\mathbb{Z}}+ \frac{\Phi}{2\pi}, \quad\quad\quad\quad\quad S^z  \in 4 {\mathbb{Z}},  \nn\\
\tilde{W} &\in&  {\mathbb{Z}}-\frac{1}{4}+ \frac{\Phi}{2\pi}, \,\quad\quad\quad S^z  \in 4 {\mathbb{Z}} +  2 ,\nn\\\eea
with $\tilde{W}$ - the winding number of $\tilde{\phi}$:
\beq \tilde{W} = \frac{1}{2\pi} (\tilde{\phi}(x+L) - \tilde{\phi}(x)).\eeq
The operator $\tau^z_{j,j+1}$ that flips the flux $e^{i \Phi}$, thus, twists the boundary condition of $\tilde{\phi}$ by $\pi$, i.e.
\beq \tau^z_{j,j+1} \sim \cos (\tilde{\theta}/2).\eeq
We note that, $S^+_j \sim e^{i \tilde{\phi}}$, is, strictly speaking, not a gauge invariant operator under the Gausses law. On the other hand, $(S^+_j)^2 \sim e^{2i \tilde{\phi}}$, is gauge invariant. It is, thus, more convenient to work with
\beq \phi = 2 \tilde{\phi}, \quad \theta = \tilde{\theta}/2.\eeq
In terms of these variables
 \beq L = \frac{1}{2\pi} \d_x \phi \d_t \theta - \frac{1}{2\pi} \left(K (\d_x \phi)^2 + \frac{1}{4 K} (\d_x {\theta})^2\right) \label{Lthetaphi},\eeq
with $K = \tilde{K}/4$ and
\bea W &\in&  {\mathbb{Z}}, \quad\quad\quad\quad\quad N \in 2 {\mathbb{Z}},  \nn\\
W &\in&  {\mathbb{Z}}+\frac{1}{2}, \,\quad\quad\quad N \in 2 {\mathbb{Z}} +  1,  \label{WNrel}\eea
where
\beq W = \frac{1}{2\pi} (\phi(x+L) - \phi(x)), \quad N = \frac{1}{2\pi}(\theta(x+L)-\theta(x)). \label{WN}\eeq
Note that $N = \frac{S^z}{2}$ is the total physical fermion number. As we recall below, the boundary conditions satisfied by $\theta$ and $\phi$ are the standard boundary conditions obtained in bosonization of a fermion theory. Further, for even $N$,  $e^{i \Phi} = (-1)^W$ and for $N$ - odd, $e^{i \Phi} = e^{\pi i (W+\frac12)}$. Therefore, we find that the $\mathbb{Z}_2$ symmetry $U$, Eq.~(\ref{US}), is given by
\beq U = (-1)^{N/2+W}.\eeq 
Note that $N/2 + W$ is an integer in both even and odd $N$ sectors.  Finally, under time-reversal ${\cal T}_{NK}$ (\ref{TNKS}), 
\beq {\cal T}_{NK}:\,\, \phi \to -\phi, \quad \theta \to \theta, \quad i \to -i. \label{T0phitheta}\eeq

Let us now make the connection with the QSHI edge theory (\ref{psiedge}) explicit. If we bosonize  (\ref{psiedge}) using $\psi_{R/L} \sim e^{-i \phi_{R/L}}$, 
\beq L = \frac{-1}{4\pi} \d_x \phi_R \d_t \phi_R + \frac{1}{4\pi} \d_x \phi_L \d_t \phi_L -\frac{1}{8\pi} (K+ K^{-1}) ((\d_x \phi_L)^2 + (\d_x \phi_R)^2) - \frac{1}{4\pi}(K-K^{-1}) \d_x \phi_R \d_x \phi_L,\eeq
where in the free fermion theory $K = 1$. The fermion densities are given by
\beq \psi^{\dagger}_R \psi_R = -\frac{1}{2\pi} \d_x \phi_R, \quad \psi^{\dagger}_L \psi_L = \frac{1}{2\pi} \d_x \phi_L.\eeq
Defining
\beq \phi = \frac{\phi_R + \phi_L}{2}, \quad \theta = \phi_L - \phi_R, \eeq
we recover the Lagrangian (\ref{Lthetaphi}). Further, the winding numbers (\ref{WN}) become
\beq N = N_L + N_R, \quad W = \frac{1}{2} (N_L - N_R).\eeq
Since (with NS boundary conditions) $N_L$ and $N_R$ are integers, $N$ and $W$ satisfy precisely the relations (\ref{WNrel}). Further, the $\mathbb{Z}_2$ symmetry $U = (-1)^{N/2+W} = (-1)^{N_L}$, in agreement with the transformation properties (\ref{Z2psi}). Also, the non-Kramers time-reversal symmetry $\TNK: \psi_R \leftrightarrow \psi_L$, $\phi_R \to - \phi_L$, $\phi_L \to -\phi_R$ gives precisely Eq.~(\ref{T0phitheta}). 


\subsection{Anomaly cocycle}
\label{sec:cocmain}

As with any non-onsite symmetry in 1d implemented by a finite depth circuit, one may extract the ``anomaly cocycle" characterizing it that matches the algebraic data of the bulk SPT.\cite{CZX,ElseNayak} We now do this for our construction in section \ref{sec:symmact} using the approach of Else and Nayak, Ref.~\onlinecite{ElseNayak}. Let us recall the steps. First, we are dealing with a fermion system with a full symmetry group $G_f$ that includes $\mathbb{Z}^f_2$ as a central subgroup. We may form the bosonic symmetry group $G_b = G_f/\mathbb{Z}^f_2$. For every element $g \in G_b$, pick a lift $\tilde{g}$ to $G_f$. Then,
\beq \tilde{g} \cdot \tilde{h} = \left((-1)^{F}\right)^{\lambda(g,h)} \,\, \widetilde{gh}, \label{lambdadef}\eeq
where $\lambda(g,h) \in \{0,1\}$. In fact, $\lambda$ is a two-cocycle: 
\beq (d \lambda)(g,h,k) = \lambda(h,k) - \lambda(gh,k) + \lambda(g,hk)-\lambda(g,h) = 0 \,\,\,\, (mod \,\,2).\eeq
Thus, we can think of $\lambda$ as an element of $H^2(G_b, \mathbb{Z}_2)$, where the co-boundary transformations correspond to picking a different lift $\tilde{g}$ to $G_f$. 

Else and Nayak assume that for every element $g \in G_f$, the symmetry acts on the boundary as a finite depth unitary ${\cal U}(g)$ if $g$ is not-time reversing. In case when $g$ is a time-reversing, ${\cal U}(g)$ is a finite depth unitary times the anti-unitary operator ${\cal T}_0$ in Eq.~(\ref{T0}).  This, in particular, means that we can truncate ${\cal U}(g)$ to a finite region $[a,b]$ of the boundary. Let us call this truncation ${\cal U}^r(g)$. (In case of time-reversing $g$, ${\cal U}^r(g)$ is a finite depth unitary acting on $[a,b]$ times ${\cal T}_0$.) We require the truncation ${\cal U}^r(g)$ to be fermion parity even. Then,
\beq {\cal U}^r(\tilde{g}) {\cal U}^r(\tilde{h}) = L(g,h) R(g,h) \Pi^{\lambda(g,h)} {\cal U}^r(\widetilde{gh}), \quad g, h \in G_b, \label{Lghdef}\eeq
Here $\Pi$ is the restriction of $(-1)^F$ to the interval $[a,b]$, and $L(g,h)$ and $R(g,h)$ are unitaries acting near left and right boundaries of $[a,b]$.  We introduce a cochain $\sigma(g,h) \in \{0,1\}$: if $L(g,h)$ and $R(g,h)$ are bosonic then $\sigma(g,h) = 0$, if they are fermionic then $\sigma(g,h) = 1$. Associativity requires
\beq L(g,h)L(gh, k) = w_3(g,h,k) \,\,^gL(h,k) L(g,hk), \quad g, h, k \in G_b, \label{w3def}\eeq
where $^gL(h,k) = {\cal U}^r(\tilde{g}) L(h,k) ({\cal U}^r(\tilde{g}))^{-1}$ and $w_3(g,h,k) \in U(1)$. We see that this implies $d \sigma = 0 \,\, (mod\,\,2)$. In fact, $\sigma(g,h) \in H^2(G_b,\mathbb{Z}_2)$, where the co-boundary transformation corresponds to the freedom of using different truncations ${\cal U}^r(g)$. Furthermore, $w_3$ satisfies
\beq d_T w_3 = (-1)^{\sigma \cup \sigma + \lambda \cup \sigma} \label{dTw3}. \eeq
Here, $(d_T w_3)(g,h,k,l) = \, ^gw_3(h,k,l) w^{-1}(gh,k,l) w(g,hk,l) w^{-1}(g,h,kl) w(g,h,k)$, and the action $^g u = u$ if $g$ is not time-reversing and $^g u = u^*$ if $g$ is time-reversing. For cochains $a \in C^n(G_b, \mathbb{Z}_2)$, $b \in C^m(G_b, \mathbb{Z}_2)$, $a\cup b \in C^{n+m}(G_b,\mathbb{Z}_2)$ is defined to be $(a\cup b)(g_1,g_2, \ldots, g_{n+m}) = a(g_1, g_2, \ldots, g_n) b(g_{n+1}, g_{n+2}, \ldots g_{n+m})$. 

Thus, the non-onsite symmetry is characterized by two pieces of  algebraic data $\sigma \in H^2(G_b, \mathbb{Z}_2)$ and $w_3 \in C^3(G_b, U(1))$, subject to Eq.~(\ref{dTw3}). Under ``gauge transformations":
\bea \sigma &\to& \sigma + d \alpha, \quad \lambda \to \lambda + d \beta, \nn\\
w_3 &\to& w_3 (-1)^{\alpha \cup (\sigma + d\alpha) + (\lambda + \sigma) \cup \alpha + \beta \cup (\sigma + d \alpha)} (d_T \gamma) \label{gaugecoc}\eea
for $\alpha, \beta \in C^1(G_b, \mathbb{Z}_2)$, $\gamma \in C^2(G_b, U(1))$, 
and   $(d \alpha)(g,h) = \alpha(g) + \alpha(h) - \alpha(g h)$, 
\beq (d_T \gamma)(g,h,k) = 
\,\,^g\gamma(h,k) \gamma^{-1}(gh,k) \gamma(g,hk) \gamma^{-1}(g,h).\eeq

Let's now apply this procedure to our case. We have $G_b =  \mathbb{Z}_2 \times O(2) $, where $O(2)$ combines the $U(1)$ symmetry and time-reversal ${\cal T}_{NK}$, and $\mathbb{Z}_2$ is the unitary symmetry $U$, Eq.~(\ref{Udef}). We parameterize  $g \in G_b$ by three elements $g_1, g_2 \in \{0,1\}$ and $g_3 \in [0,\pi)$. The group operation in $G_b$ is
\beq (g_1, g_2, g_3)\cdot (h_1, h_2, h_3) = (g_1 + g_2, g_2 + h_2, \{g_3 + (-1)^{g_2} h_3\}).\eeq
Here and below for $x \in [l \pi, (l+1)\pi)$, we define $\{x\} = x - l \pi \in [0, \pi)$. We choose the following lifts to $G_f$:
\beq \tilde{g} = U^{g_1} e^{i g_3 N} (\TNK)^{g_2}.\eeq 
This gives 
\beq \lambda(g, h) = g_2 h_1 + \langle g_3 + (-1)^{g_2} h_3\rangle \quad (mod\,\,2). \label{lambda}\eeq
Here and below $\langle x \rangle = \frac{1}{\pi} (x - \{x\}) \in \mathbb{Z}$. Performing the restriction of symmetry operators to a finite interval and computing $L(g,h)$, we obtain
\beq \sigma(g,h) = g_1 h_1, \quad\quad w_3(g,h,k) = i^{g_1 h_1 k_1 + \langle g_3 + (-1)^{g_2} h_3 \rangle k_1}. \label{w3final}\eeq
We sketch the details of the computation in appendix \ref{app:restrict}. It is easy to check that $\lambda, \sigma, w_3$ above, indeed, satisfy Eq.~(\ref{dTw3}). 

Before concluding this section, we would like to point out appendix \ref{app:Ugen}, where we essentially reverse the procedure above. For any finite $G_f$, given $\sigma$ and $w_3$, we construct a non-onsite symmetry characterized by this data.

\section{Bulk + boundary construction}
\label{sec:constr}
In this section, we present a commuting projector Hamiltonian for the bulk of the fermion SPT with just $U(1)$ and unitary $\mathbb{Z}_2$ symmetry (\ref{Z2psi}).   Our construction essentially amounts to two copies of the Tarantino-Fidkowski (TF) model\cite{FidkowskiSpin} for the $\mathbb{Z}_2 \times \mathbb{Z}^f_2$ SPT. We then derive an effective 1d lattice model for the edge and show that it matches the bosonized model in section \ref{sec:JW}. For completeness, we review here many details of Refs.~\onlinecite{FidkowskiSpin, MJ}.

\subsection{Bulk}

\begin{figure}[h!]
\centering
\includegraphics[width=0.8\textwidth]{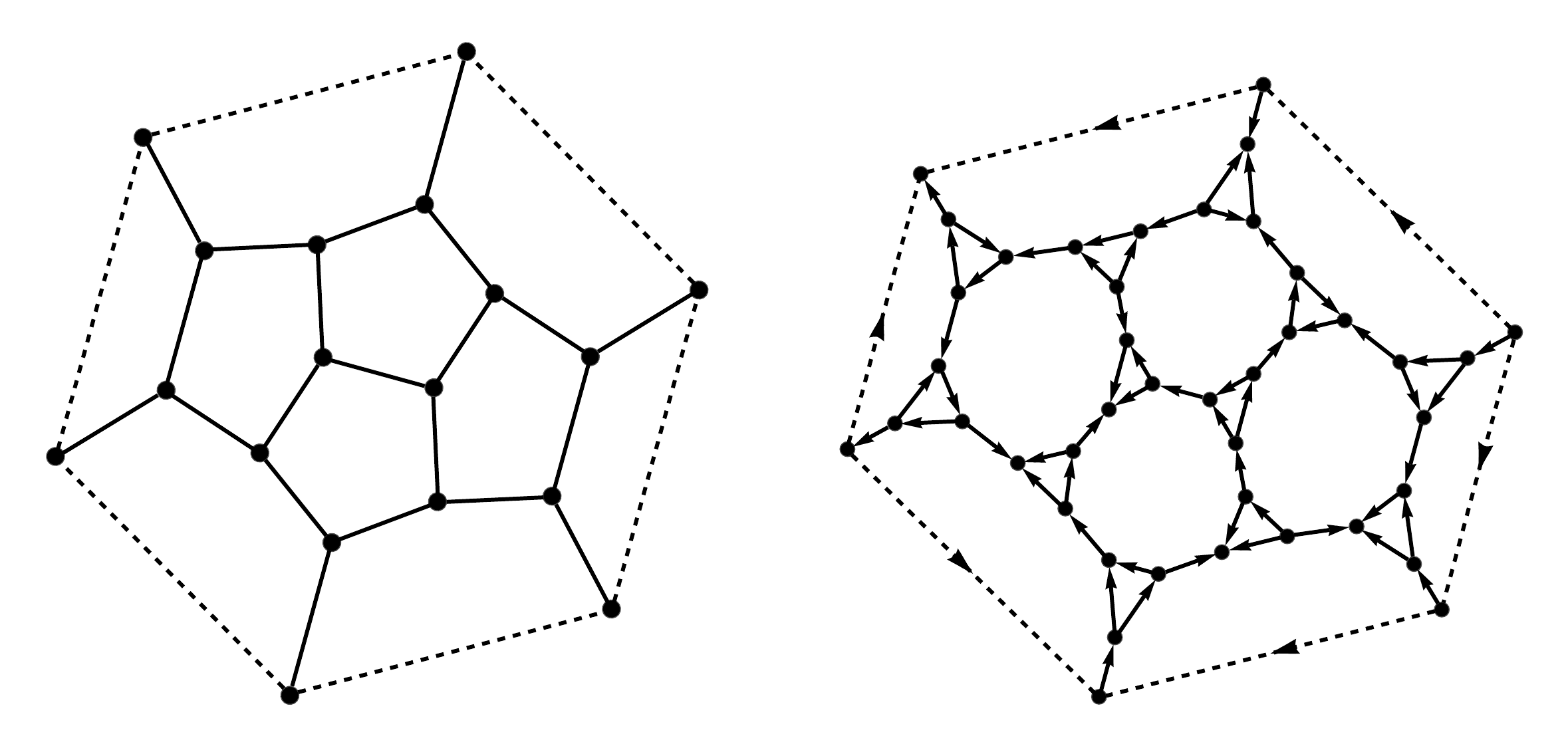}
\caption{Bulk+boundary construction. Left: graph \({\cal G}\) which illustrates ``physical'' plaquettes; Ising spins live on the faces of this graph. Boundary edges are dashed. Right: graph \({\cal G}'\) where the vertices are complex fermions. ${\cal G}'$ is given a Kasteleyn orientation: all edges are oriented so that there is an odd number of clockwise edges around any face.
}
\label{fig:Gopen}
\end{figure}

 We consider a trivalent graph ${\cal G}$ embedded into a closed oriented surface (see Fig.~\ref{fig:Gopen}, left, bulk). We blow up each vertex of this trivalent graph into a triangle obtaining a new trivalent graph ${\cal G}'$ (see Fig.~\ref{fig:Gopen}, right, bulk). We place a complex fermion $c_i$ on each vertex of ${\cal G}'$. It will be convenient to break $c_i$ up into Majorana fermions $\gamma_i$, $\bar{\gamma}_i$ as follows:
\beq c_i = \frac{1}{2} (\gamma_i + i \bar{\gamma}_i), \quad c^{\dagger}_i = \frac{1}{2} (\gamma_i - i \bar{\gamma}_i).\eeq
We let the fermion number $N = \sum_i c^{\dagger}_i c_i$ - this generates the $U(1)$ symmetry of the model.
 
 The edges of ${\cal G}'$ are divided into two groups: type I edges connecting vertices belonging to different triangles and type II edges connecting vertices within the same triangle. 
 The faces of ${\cal G}'$ are also divided into two groups: the faces inherited from the faces of ${\cal G}$ - we call these faces plaquettes,  and new triangular faces coming from the blown up vertices. We place a dynamical Ising spin variable $\tau^z_p$ on each plaquette $p$.  The $\mathbb{Z}_2$ symmetry of the model acts by flipping all the plaquette spins,
 \beq U = \prod_p \tau^x_p. \label{Ubulk}\eeq
 Note that the fermions $c_i$ are neutral under $\mathbb{Z}_2$. 
 
\begin{figure}[t]
\centering
\includegraphics[width=\textwidth]{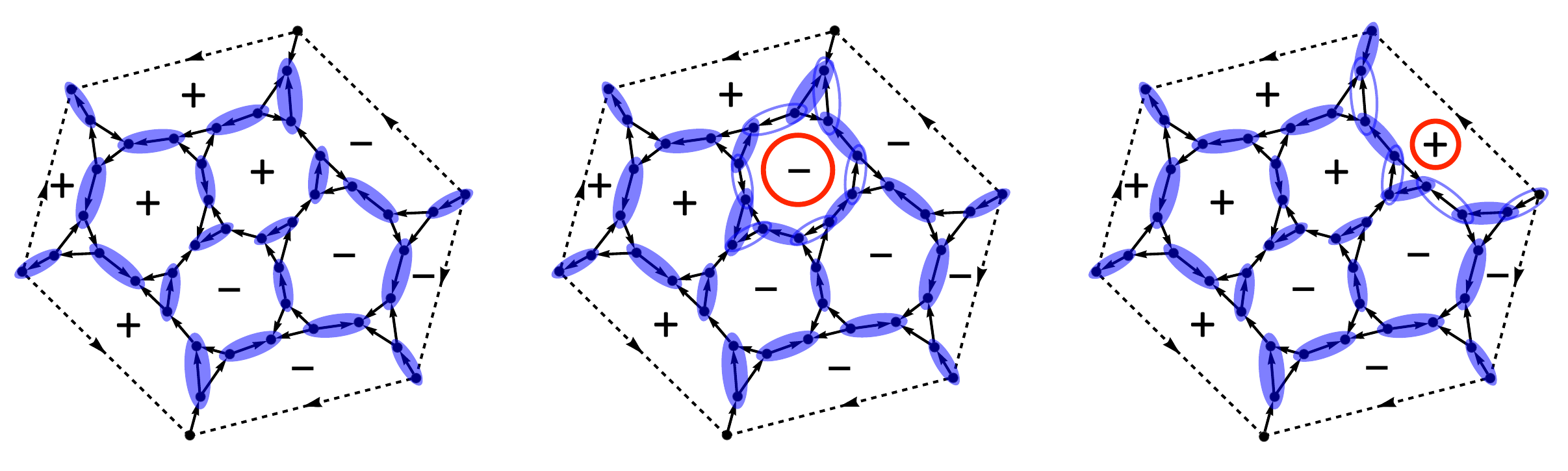}
\caption{Effect of the plaquette flip operators $F_p$. Left: Spin configuration and associated dimer cover ${\cal D}$ (filled ovals). Middle: left configuration with the bulk plaquette in red flipped  - the new dimer cover ${\cal D'}$ is shown with filled ovals.  Part of the old dimer cover $({\cal D} + {\cal D'})\cap{\cal D}$ is shown  with empty ovals. Filled and empty ovals form a closed loop $({\cal D} + {\cal D'})$ around the flipped plaquette. Right: Left configuration with boundary plaquette in red flipped. Same notation as in the middle figure. Now filled and empty ovals $({\cal D} + {\cal D'})$ form an open segment around the flipped plaquette.
} 
\label{fig:Flip}
\end{figure}

For each configuration of plaquette spins we may extend it to the triangular faces of ${\cal G}'$ by the ``majority rule": if the spin of the majority of three plaquettes bordering a triangle is $\tau$ then the triangle also gets assigned the spin $\tau$. 
Every spin configuration gives rise to a dimer covering of ${\cal G}'$: a type I edge is covered if the spins neighbouring it are the same, while a type II edge is covered if the spins neighbouring it are different (see Fig.~\ref{fig:Flip}, left).
 
We endow ${\cal G}'$ with a Kasteleyn orientation:  we specify a direction for each edge of ${\cal G}'$ such that the number of clockwise oriented edges surrounding any face is odd. This applies to both the plaquette faces and the triangles. For an edge $(ij)$, we let $s_{ij} = 1$ if the orientation points from $i$ to $j$, and $s_{ij} = -1$ otherwise. 

The Hamiltonian consists of two terms. The first term $H_{fermion}$ locks the fermion configuration to the spin configuration as follows: if the edge $(ij)$ is occupied by a dimer then $i s_{ij} \gamma_i \gamma_j  \sim 1$ and $i s_{ij} \bar{\gamma}_i \bar{\gamma}_j \sim 1$.  Explicitly:

\beq H_{fermion} = - \sum_{ \langle ij \rangle \in {\rm Type\, I} } (1-D_{ij}) i s_{ij} (\gamma_i \gamma_j + \bar{\gamma}_i \bar{\gamma}_j) - \sum_{ \langle i j\rangle \in {\rm Type\, II}} D_{ij} i s_{ij} (\gamma_i \gamma_j + \bar{\gamma}_i \bar{\gamma}_j), \label{Hferm}\eeq
where $D_{ij} = 1$ if the plaquettes bordering edge $(ij)$ have opposite spin and $D_{ij} = 0$ if plaquettes bordering edge $(ij)$ have the same spin. We note that $i (\gamma_i \gamma_j + \bar{\gamma}_i \bar{\gamma}_j) = 2i (c^{\dagger}_i c_j - c^{\dagger}_j c_i)$ so $H_{fermion}$ preserves the $U(1)$ symmetry. For a fixed spin configuration the ground state of $H_{fermion}$ is as follows: $|\psi\rangle = \prod_{(ij) \in {\cal D}} \frac{1}{\sqrt{2}} (c^{\dagger}_i - i s_{ij} c^{\dagger}_j)|0\rangle$. Here $|0\rangle$ is the Fock vacuum and ${\cal D}$ denotes the dimer configuration. Thus, there is exactly one fermion on each dimer and its wavefunction is a linear superposition of the two sites on the dimer. We call the ground state subspace of $H_{fermion}$ ${\cal V}^c$.

We next define the ``plaquette" flip operator $F_p$. Let $c$ be a spin configuration and $c_p$ be the same configuration with the spin on plaquette $p$ flipped. Let ${\cal D}(c)$ be the dimer covering associated with $p$ and ${\cal D}(c_p)$ the dimer covering associated with $c_p$. For any two dimer coverings ${\cal D}$ and ${\cal D}'$ let ${\cal D}+{\cal D}'$ consist of edges covered by ${\cal D}$ or ${\cal D}'$ but not both. Then ${\cal D}(c) + {\cal D}(c_p)$ is a closed loop whose edges belong to the plaquette $p$ or to the triangles bordering  $p$ (see Fig.~\ref{fig:Flip},  middle). Let's order vertices in this loop in a counter-clockwise fashion as $1, 2, \ldots 2n$, with edges $(12)$, $(34)$, \ldots,  $(2n-1,2n)$ belonging to ${\cal D}(c)$ and $(23)$, $(45)$, \ldots, $(2n, 1)$ belonging to ${\cal D}(c_p)$. $F_p$ moves fermions from the former to the latter set of edges (see Fig.~\ref{fig:create}, top). More explicitly,
 \beq F_p=\sum_c X_{p,c}\otimes(\tau_p^x P_{p,c}), \label{Fpdef} \eeq
Here,  $c$ runs over the spin configurations of $p$ and plaquettes neighbouring $p$. $P_{p,c}$ is a projector on the corresponding spin configuration. Thus, the term in brackets only acts on the spin degrees of freedom, selecting the configuration $c$ and then flipping the spin on plaquette $p$. On the other hand, $X_{p,c}$ acts on the fermions via

\beq X_{p,c} = 2^{n-1} \prod_{i = 1}^{n} (P_{2i, 2i+1} \bar{P}_{2i,2i+1})  \prod_{i=1}^{n}(P_{2i-1,2i}\bar{P}_{2i-1,2i}), \eeq
where $P_{ij} = \frac{1}{2}(1+ i s_{ij} \gamma_i \gamma_j)$ and $\bar{P}_{ij} = \frac{1}{2}(1+is_{ij} \bar{\gamma}_i \bar{\gamma}_j)$ are projectors. The second product above projects the Majorana fermions in ${\cal D}(c)+ {\cal D}(c_p)$ onto the initial dimer cover ${\cal D}(c)$, and the first product - onto the final dimer cover ${\cal D}(c_p)$. Note that the second product acts trivially on states in  ${\cal V}^c$. Also $X_{p,c}$ is independent of the base-point chosen for enumerating the vertices in ${\cal D}(c) + {\cal D}(c_p)$. Further, $F_p$ thus defined is Hermitian and preserves $\mathbb{Z}_2$ and $U(1)$ symmetries. To see the last point, note that
\beq P_{ij} \bar{P}_{ij} = \frac{1}{4} (1+ i s_{ij} \gamma_i \gamma_j)(1+is_{ij} \bar{\gamma}_i \bar{\gamma}_j) = \frac{1}{4}\left(1 + 2 i s_{ij} (c^{\dagger}_i c_j - c^{\dagger}_j c_i) - 4 (c^{\dagger}_i c_i - \frac12)(c^{\dagger}_j c_j -\frac12) \right),\eeq
which  explicitly preserves the fermion number $N$. Finally, $F_p$ preserves the subspace ${\cal V}^c$.

We recall that the TF model is almost identical, except it has only one set of Majorana fermions $\gamma_i$ and the operator $X_{p,c}$ in the plaquette term $F^{TF}_p$ acts as
\beq X^{TF}_{p,c} = 2^{(n-1)/2} \prod_{i = 1}^{n} P_{2i, 2i+1}  \prod_{i=1}^{n}P_{2i-1,2i}. \eeq
Thus, we can directly import many properties of the TF model. In particular, in the subspace ${\cal V}^c$,  $[F_p, F_q] = 0$ for any two plaquettes $p$ and $q$, and $F^2_p = 1$.


We take the full Hamiltonian to be 
\beq H_{bulk} = H_{fermion} - \sum_p F_p.\eeq
The ground state is unique: $|\psi\rangle = \prod_p (1+ F_p) |+\rangle$, where $|+\rangle$ is the state with all spins up and fermion slaved to dimers accordingly. 


\subsection{Introducing the edge}

\begin{figure}[h!]

\includegraphics[width=0.9\textwidth]{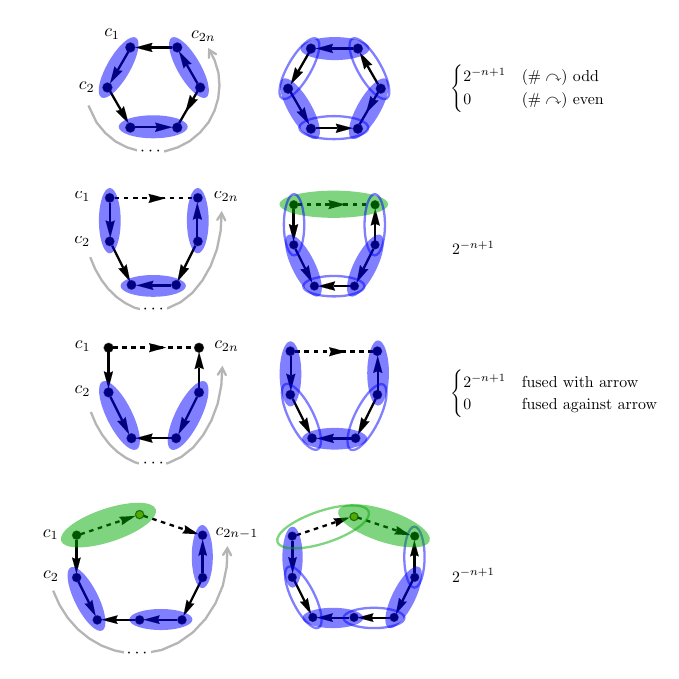}
\caption{Plaquette flip operator $F_p$: the fermion factor $X_{p,c}$. The left and right column show the initial and final configurations. Filled and empty blue ovals indicate dimers as in Fig.~\ref{fig:Flip}. The top row corresponds to bulk plaquette flips, and the rest to boundary plaquette flips.  For each row, we suppose the system is initially in the state on the left, we then apply the projectors $P_{ij} \bar{P}_{ij}$ corresponding to the blue filled ovals on the right. \newline
First row:   rotating dimers around a closed loop annihilates the state if the loop is not Kasteleyn oriented, and otherwise shrinks its norm to $2^{-n+1}$.\newline
Second row:  A solid green dimer indicates the fusion channel of unpaired fermions: $i s_{1, 2n} \gamma_1 \gamma_{2n} = 1$,  $i s_{1, 2n} \bar{\gamma}_1 \bar{\gamma}_{2n} = 1$ 
Thus, the two unpaired fermions are fused to respect the Kasteleyn orientation and the norm is  \(2^{-n+1}\). \newline
Third row: If the fusion state of the unpaired fermions on the left is $i s_{1, 2n} \gamma_1 \gamma_{2n}  = -1$ or $i s_{1, 2n} \bar{\gamma}_1 \bar{\gamma}_{2n} = -1$, the state is annihilated, otherwise, the norm is \(2^{-n+1}\).
\newline
Fourth row: There are $2n-1$ fermions in ${\cal D}(c)+{\cal D}(c_p)$. The gray  fermion on top is auxilliary (e.g. another unpaired boundary fermion from a different plaquette) and is assumed to be initially fused with $c_1$ along the green oval. In the final state, it becomes fused with $c_{2n-1}$.}

\label{fig:create}
\end{figure}

We can likewise adapt the discussion of the edge of the TF model in Ref.~\onlinecite{MJ} to the present case. As in Ref.~\onlinecite{MJ} we start with a trivalent graph ${\cal G}$ with a boundary  (see Fig.~\ref{fig:Gopen}, left) and blow up all the vertices, except the boundary vertices, into triangles, obtaining a new graph ${\cal G}'$  (see Fig.~\ref{fig:Gopen}, right). We place a complex fermion $c_i$ at each vertex of ${\cal G}'$. We also choose a Kasteleyn orientation on ${\cal G}'$. As before, the faces of ${\cal G}'$ inherited from ${\cal G}$ are called plaquettes. Each plaquette (including boundary plaquettes) carries an Ising spin, and the spin configuration can be extended to the ``triangles" via the majority rule. 
Each spin configuration gives rise to a dimer configuration as before: the only difference is that boundary edges never carry dimers (we can think of them as type III). As before, we lock fermions to the dimer configurations by the Hamiltonian $H_{fermion}$ (\ref{Hferm}), whose ground state subspace we denote as ${\cal V}^c$. Note that every domain wall on the boundary gives rise to an ``unpaired" complex fermion $c_i$ on the boundary vertex.

We take the Hamiltonian to be
\beq H = H_{fermion} - \sum_{p \in bulk} F_p\eeq
Clearly we have ground states for every configuration of boundary spins $\tau$. 
Furthermore, for a fixed configuration of boundary spins we have a $2^{N_{dw}}$ degeneracy associated with the unpaired complex boundary fermions. Here $N_{dw}$ is the number of boundary domain walls. Thus, if we have $L$ boundary plaquettes, the ground space subspace associated with the boundary degeneracy has dimension 
\beq g = 2 \sum_{N_{dw} -even}^{L} \left(\begin{array}{c} L\\N_{dw}\end{array}\right) 2^{N_{dw}} = 3^L + (-1)^L. \eeq
Our goal will be to find a convenient basis for this subspace.

We introduce the boundary plaquette flip operators $F_p$ analogously to Ref.~\onlinecite{MJ}. We again use the form (\ref{Fpdef}), but modify $X_{p,c}$ as follows. There are several cases to consider:

\begin{itemize}
\item Both boundary fermions of $p$ in $c$ are paired  (Fig.~\ref{fig:create}, second row, left). Then after acting with $F_p$ both fermions will be unpaired  (Fig.~\ref{fig:create}, second row, right).  ${\cal D}(c) + {\cal D}(c_p)$ is an open string containing $2n$ fermions, which we label consecutively along the string so that $1$ and $2n$ are the boundary fermions.  We let
\beq X_{p,c}=N_{p,c} \prod_{i = 1}^{n-1} (P_{2i, 2i+1} \bar{P}_{2i, 2i+1}), 
\label{Xbound1} \eeq
with $N_{p,c} = 2^{n-1}$. For $|\psi\rangle \in {\cal V}^c$, $F_p |\psi\rangle$ has the same norm as $|\psi\rangle$. Further, $i s_{1,2n} \gamma_{1} \gamma_{2n}  \sim i s_{1,2n} \bar{\gamma}_{1} \bar{\gamma}_{2n} \sim 1$ on $F_p |\psi\rangle$, where $s_{1,2n}$ corresponds to the orientation of the boundary edge $(1,2n)$.

\item Both boundary fermions of $p$ in $c$ are unpaired, Fig.~\ref{fig:create}, third row, left. Then after acting with $F_p$ both fermions will be paired, Fig.~\ref{fig:create}, third row, right. Again, ${\cal D}(c) + {\cal D}(c_p)$ is an open string with $2n$ fermions, $1$ and $2n$ being the boundary fermions. We let
\beq X_{p,c}=N_{p,c}  \prod_{i =1}^n (P_{2i-1,2i} \bar{P}_{2i-1,2i}),
\label{Xbound2} \eeq
with $N_{p,c} = 2^{n-1}$. Now, $F_p |\psi\rangle$ has the same norm as $|\psi\rangle$ if $i s_{1,2n} \gamma_1 \gamma_{2n} |\psi \rangle = |\psi\rangle$ and $i s_{1,2n} \bar{\gamma}_1 \bar{\gamma}_{2n} |\psi \rangle = |\psi\rangle$.  If $i s_{1,2n} \gamma_1 \gamma_{2n} |\psi \rangle = - |\psi\rangle$ or $i s_{1,2n} \bar{\gamma}_1 \bar{\gamma}_{2n} |\psi \rangle = - |\psi\rangle$then $F_p |\psi \rangle  =0$. 

\item One boundary fermion  of $p$ in $c$ is paired and the other is unpaired, Fig.~\ref{fig:create}, fourth row, left. Then ${\cal D}(c) + {\cal D}(c_p)$ is an open string containing $2n-1$ fermions, which we label consecutively. We let $1$ be the initially unpaired fermion and $2n- 1$ be the initially paired fermion. After the flip, $1$ is paired and $2n-1$ is unpaired, Fig.~\ref{fig:create}, fourth row, right. Then
\beq X_{p,c}=N_{p,c}  \prod_{i = 1}^{n-1} \left(P_{2i-1,2i} \bar{P}_{2i-1,2i}\right), 
\label{Xbound3} \eeq
with $N_{p,c} = 2^{n-1}$. Again, $F_{p} |\psi\rangle$ has the same norm as $|\psi\rangle$. 

\end{itemize}

 Let us label boundary vertices by $i  =1, 2, \ldots, L$ and boundary plaquettes by corresponding boundary edges $(i,i+1)$. The plaquette flip operators thus defined have the following properties, which follow from Ref.~\onlinecite{MJ}:

\begin{enumerate}

\item The boundary plaquette operators preserve ${\cal V}^c$, as well as $U(1)$ and $\mathbb{Z}_2$ symmetry.

\item
The boundary \(F_p\)'s all commute with all the bulk \(F_p\)'s. Nearest neighbor boundary \(F_p\)'s do not commute, but otherwise boundary \(F_p\)'s do.

\item
Let $(i,i+1)$ be a boundary plaquette, and $c_i$, $c_{i+1}$ - its boundary fermions. If both $c_i$ and $c_{i+1}$ are unpaired
in $|\psi\rangle$, $F^2_{i,i+1} |\psi\rangle = \frac14 (1 + i s_{i,i+1} \gamma_{i} \gamma_{i+1}) (1 + i s_{i,i+1} \bar{\gamma}_{i} \bar{\gamma}_{i+1}) |\psi\rangle$. Otherwise, $F^2_i |\psi\rangle = |\psi\rangle$.

\item 
Consider adjacent boundary plaquettes $(i-1,i)$, $(i,i+1)$, and  a state $|\psi\rangle$ where the boundary fermion $c_i$ shared by these plaquettes is paired (i.e. plaquettes $(i-1,i)$ and $(i,i+1)$ have the same spin). Then,
\begin{itemize} 
\item If the   boundary fermions $c_{i-1}$ and $c_{i+1}$ are paired, $[F_{i-1,i}, F_{i,i+1}] |\psi \rangle = 0$. 
\item If the  boundary fermions  $c_{i-1}$  and $c_{i+1}$ are unpaired, $[F_i, F_{i+1}] |\psi \rangle = 0$.
\item If the  boundary fermion $c_{i-1}$ is unpaired, and the  boundary fermion $c_{i+1}$ is paired, $F_{i-1,i} F_{i,i+1} |\psi \rangle = \frac{1}{2} F_{i,i+1} F_{i-1,i} |\psi\rangle$. 
\item If the  boundary Majorana $c_{i-1}$  is paired and the   boundary Majorana $c_{i+1}$  is unpaired, $F_{i,i+1} F_{i-1,i} |\psi \rangle = \frac{1}{2} F_{i-1,i} F_{i, i+1} |\psi\rangle$. 

\end{itemize}

\item
Let $(i,i+1)$ be a boundary plaquette, and  $|\psi\rangle$ a state where $c_{i}$ is unpaired but $c_{i+1}$ is paired. Then,  $F_{i,i+1}\gamma_{i} | \psi\rangle= s_{i,i+1} \gamma_{i+1} F_i  |\psi\rangle \). 
Likewise, if $c_{i}$ is paired but $c_{i+1}$ is unpaired then $F_{i,i+1} \gamma_{i+1} |\psi\rangle = s_{i,i+1} \gamma_{i} F_{i,i+1} |\psi\rangle$. Similarly for $\gamma \to \bar{\gamma}$. 

\end{enumerate}

\subsection{Labeling the edge Hilbert space}

We now introduce a convenient way to label the states in the edge Hilbert space.  We assume that the bulk has a disk topology so that the boundary is a circle. In our construction, this effectively corresponds to Neveu-Schwarz boundary conditions on the boundary fermions: the Kasteleyn orientation satisfies $\prod_{i=1}^{L} s_{i,i+1} = -1$. 

To label a state we need to specify the spins of the boundary plaquettes, as well as the state of the unpaired boundary fermions. Let $|+\rangle$ be the state where all the boundary spins are ``+" and so there are no unpaired fermions.  We can build a state where the consecutive ``$-$" domains  are between sites $(i_1, j_1)$, $(i_2, j_2)$, \ldots, $(i_{N_d}, j_{N_d})$:
\beq |\psi\rangle = \prod_{l = 1}^{N_d} F_{(i_l, j_l)} |+\rangle, \label{psidom} \eeq
where $N_d$ is the number of ``$-$" domains and 
\beq F_{(i,j)} = F_{j-1,j} F_{j-1,j-2} \ldots F_{i+1,i+2} F_{i,i+1}.\eeq
Note that $F_{(i_l, j_l)}$ for different $l$ commute since the domains do not touch. Using the properties of plaquette operators, $|\psi\rangle$ in (\ref{psidom}) is a normalized state with $i s_{(i_l, j_l)} \gamma_{i_l} \gamma_{j_l} \sim 1$ and  $i s_{(i_l, j_l)} \bar{\gamma}_{i_l} \bar{\gamma}_{j_l} \sim 1$ on $|\psi\rangle$. Here, $s_{(i,j)} = s_{j-1, j} s_{j-1,j-2} \ldots s_{i+1,i+2} s_{i,i+1}$. Eq.~(\ref{psidom}) is just one state in the  fermion Hilbert space corresponding to the fixed boundary spin configuration. Further, it is more convenient to work with the basis of states where the fermion occupation number $n_i = c^{\dagger}_{i} c_i = \frac12 (1 + i \gamma_i \bar{\gamma}_i)$ of each unpaired fermion is specified. Suppose we want to build a state with the domain structure as above and occupation numbers $n_{i_l}, n_{j_l} \in \{0,1\}$. For a domain $(i_l, j_l)$ we can do this by acting with operators 
\beq B_{i_l}(n_{i_l}, n_{j_l}) = \left\{\begin{array}{c} \sqrt{2} c^{\dagger}_{i_l} c_{i_l}, \quad\quad n_{i_l} = 1, n_{j_l} = 0 \\
\sqrt{2} c_{i_l} c^{\dagger}_{i_l}, \quad\quad n_{i_l} = 0, n_{j_l} = 1\\
\sqrt{2} s_{(b,i_l)} c_{i_l},  \quad\quad n_{i_l} = 0, n_{j_l} = 0\\
\sqrt{2} s_{(b,i_l)} c^{\dagger}_{i_l,}  \quad\quad n_{i_l} = 1, n_{j_l} = 1
\end{array} \right.\eeq
on $|\psi\rangle$ in Eq.~(\ref{psidom}). Here $b$ is an arbitrary (but fixed) ``basepoint" whose purpose will become clear shortly. One can check that $B_{i_l}|\psi\rangle$ is a normalized state with the desired properties. To generalize to an arbitrary state,
\beq |\{(i_l, j_l)\}; \{n_{i_l}, n_{j_l}\}\rangle_b = \prod_{l=1}^{N_d} B_{i_l}(n_{i_l}, n_{j_l}) \prod_{l = 1}^{N_d} F_{(i_l, j_l)} |+\rangle \label{ijn},\eeq
where the operators in the first product are ordered with smaller $l$ on the left. An important technical point is that we select an order of the consecutive ``$-$" domains such that $(i_1, j_1)$ is the first ``$-$" domain fully to the right of the base-point $b$ (this includes the case when $i_1 = b$). Under a change of base-point,
\beq |\{(i_l, j_l)\}; \{n_{i_l}, n_{j_l}\}\rangle_{b+1} =  (s_{b,b+1} u)^N |\{(i_l, j_l)\}; \{n_{i_l}, n_{j_l}\}\rangle_{b}. \eeq
Here $N$ is the fermion number, which we always count relative to that of the $|+\rangle$ state. The phase   $u = -1$ if $b$ is the beginning of a $``-"$ domain $(i=b,j)$ such that $n_{i} = n_{j} = 0$ or $n_{i} = n_{j} = 1$. Otherwise, $u = 1$. Note that in the even fermion sector the states (\ref{ijn}) are independent of the basepoint, while in the odd fermion sector they transform by a phase. 

Note that the numbers $n_i, n_j \in \{0,1\}$ are defined only on sites corresponding to unpaired fermions. It is convenient to work in an enlarged Hilbert space, where we place a spin $1$ on each site $i$ of the lattice. If there is a domain wall at site $i$, i.e. $\tau^z_{i-1,i} \tau^z_{i,i+1} = -1$, we let $S^z_i = 1$ if $n_i = 1$ and $S^z_i = -1$ if $n_i = 0$. If there is no domain wall on site $i$, i.e. $\tau^z_{i-1,i} \tau^z_{i,i+1} = 1$ then $S^z_i = 0$. Thus, $S^z_i$ satisfies the constraint:
\beq (-1)^{S^z_i} = \tau^z_{i-1,i} \tau^z_{i,i+1}. \label{Gauss2}\eeq
This constraint exactly agrees with the Gausses law (\ref{Gi}) we found in section \ref{sec:unconstr}.  Note that the total fermion number $N = \frac{1}{2} S^z$, where the total spin $S^z = \sum_i S^z_i$ must always be even. We define states in this enlarged Hilbert space as follows. In the even fermion parity ($S^z = 0 \,\,(mod\,\,4)$) sector, we simply assign
\beq |\{S^z_i, \tau^z_{i,i+1} \}\rangle_b = |\{(i_l, j_l)\}; \{n_{i_l}, n_{j_l}\}\rangle_b, \quad N - even.\eeq
In the odd fermion parity sector($S^z = 2 \,\,(mod\,\,4)$)  we let
\beq |\{S^z_i, \tau^z_{i,i+1} \} \rangle_b = v |\{(i_l, j_l)\}; \{n_{i_l}, n_{j_l}\}\rangle_b, \quad N - odd,\eeq
with 
\bea v &=& 1, \quad\quad\quad\quad\,\,\, \tau^z_{b-1,b} = +1, \nn\\
v &=& i S^z_{j_{N_d}}, \quad\quad \,\,\,\,\, \tau^z_{b-1,b} = -1. \label{eq:vdef}\eea
Note that $S^z_{j_{N_d}} = \pm 1$, and if $\tau^z_{b-1,b} = -1$ and $\tau^z_{b,b+1} = 1$ then $j_{N_d} = b$, while if $\tau^z_{b-1,b} = \tau^z_{b,b+1} = -1$ then $j_{N_d}$ is just the end-point of the ``$-$" domain that crosses $b$. The reason for introducing the phase $v$ is that now we have the simple transformation law under base-point change:
\beq  |\{S^z_i, \tau^z_{i,i+1} \}\rangle_{b+1} = s_{b,b+1} e^{-i\pi S^z_b/2} |\{S^z_i, \tau^z_{i,i+1} \}\rangle_b, \quad N - odd.\eeq

\subsection{Operators}
We now describe how various operators act in the ``bosonized" Hilbert space. We begin with the plaquette operators $F_{i,i+1}$. Proceeding as in appendix B of Ref.~\onlinecite{MJ}, we find for $N$ - even,
\bea F_{i,i+1} &=& \left(\frac{1}{2}(1-\tau^z_{i-1,i} \tau^z_{i+1,i+2}) + \frac{1}{2 \sqrt{2}} (1+\tau^z_{i-1,i} \tau^z_{i+1,i+2})\right)  \frac12 (S^-_{i} \tau^x_{i,i+1} S^+_{i+1} + S^+_i \tau^x_{i,i+1}  S^-_{i+1}), \nn\\ && N - even, \label{Feven}\eea
 while for $N$ - odd, we have the same expression except when $i  = b-1$,
\bea F_{b-1,b} &=& \left(\frac{1}{2}(1-\tau^z_{b-2,b-1} \tau^z_{b,b+1}) + \frac{1}{2 \sqrt{2}} (1+\tau^z_{b-2,b-1} \tau^z_{b,b+1})\right)  \frac12 ( -i S^-_{b-1} \tau^x_{b-1,b} S^+_{b} + i S^+_{b-1} \tau^x_{b-1,b}  S^-_{b}), \nn\\ && N - odd. \label{Fodd}\eea
Note that here and below, unless otherwise specified, we assume the base-point $b$ on all our basis states $|\{S^z_i, \tau^z_{i,i+1} \} \rangle_b$. 
The  terms proportional to $\frac{1}{2}(1-\tau^z_{i-1,i} \tau^z_{i+1,i+2})$ in Eqs.~(\ref{Feven}), (\ref{Fodd}) control domain wall motion while the terms proportional to $\frac{1}{2}(1+\tau^z_{i-1,i} \tau^z_{i+1,i+2})$ control domain wall pair creation/annihilation. Note that $F_{i,i+1}$ commutes with the constraint (\ref{Gauss2}), as necessary. Also, to go between the even and odd fermion parity sector, we thread flux $\pi/2$ of $U(1)_{S^z}$ symmetry.
The form of $F_{i,i+1}$ essentially matches the Hamiltonian $H_{i,i+1}$, Eqs.~(\ref{eq:Hspin1}), (\ref{eq:Hspin2}) of section \ref{sec:unconstr}, with the basepoint $b = 1$.  (There is a $\sqrt{2}$ difference in the relative amplitude  of domain wall moving and domain wall creating terms, however, this can be adjusted without breaking the symmetry.)

Next, we discuss the form of fermion operators $c_i$, $c^{\dagger}_i$ projected into the constrained subspace ${\cal V}^c$. Proceeding as in appendix C of Ref.~\onlinecite{MJ}, we find
\bea s_{(b, i)} c_i &=& \frac{(S^-_i)^2}{2}\exp\left(\mp\frac{\pi i}{2} \sum_{j = b}^{i-1} S^z_j\right) \left(\frac{(1+\tau^z_{i-1,i})(1-\tau^z_{i,i+1})}{4} \mp  \frac{(1-\tau^z_{i-1,i})(1+\tau^z_{i,i+1})}{4}\right), \nn\\
s_{(b, i)} c^{\dagger}_i &=& \frac{(S^+_i)^2}{2}\exp\left(\mp\frac{\pi i}{2} \sum_{j = b}^{i-1} S^z_j\right) \left(\frac{(1+\tau^z_{i-1,i})(1-\tau^z_{i,i+1})}{4} \pm  \frac{(1-\tau^z_{i-1,i})(1+\tau^z_{i,i+1})}{4}\right). \nn\\ \label{cb}
\eea
Here the top and bottom signs in Eqs.~(\ref{cb}) correspond to $c_i$, $c^{\dagger}_i$ acting on states with even and odd fermion parity, respectively. It is easy to check the following anticommutation relations: $\{c_i, c_j\} = 0$, $\{c^{\dagger}_i, c_j\} = \delta_{ij} (S^z_i)^2$. Note that for $i  = j$ the last anticommutator differs from the canonical one. This is due to the fact that operators $c_i$ are first projected onto the constrained Hilbert space ${\cal V}^c$.

We note that the fermion operators (\ref{cb}) agree up to an overall phase with the JW transformed fermion operators of section \ref{sec:unconstr} (see appendix \ref{app:ferm}).

\subsection{Symmetry action}
We may derive the action of the $\mathbb{Z}_2$ symmetry $U$ in Eq.~(\ref{Ubulk}) on the bosonized states as follows. First, we note that $U|+\rangle = |-\rangle$ where $|-\rangle = F_{b-1,b}F_{b-2,b-1}\ldots F_{b+1,b+2} F_{b,b+1}|+\rangle$ is a state with all $\tau^z_{i,i+1} = -1$ and hence all $S^z_i  =0$. 

We now consider the action of $U$ on a state $|\psi\rangle$ with consecutive ``$-$" domains $(i_k, j_k)$, $k  = 1 \ldots N_d$, and corresponding occupation numbers $n_{i_k}, n_{j_k} \in \{0,1\}$ (i.e. $S^z_{i_k} = 2 n_{i_k} - 1$, $S^z_{j_k} = 2 n_{j_k} -1$). Since $U$ does not change $S^z_i$ we can choose a convenient basepoint, e.g. $b = i_1$. 
We, thus, have
\beq |\psi\rangle =  \prod_{l=1}^{N_d} B_{i_l}(n_{i_l}, n_{j_l}) \prod_{l = 1}^{N_d} F_{(i_l, j_l)} |+\rangle. \eeq
(Since  $\tau^z_{i_1-1,i_1} = +1$, there is no extra phase factor $v$, (\ref{eq:vdef}), even in the odd fermion parity sector). Acting with $U$ and recalling that the operators $c_i$ and $F_{i,i+1}$ are $\mathbb{Z}_2$ even,
\beq U|\psi\rangle = \prod_{l=1}^{N_d} B_{i_l}(n_{i_l}, n_{j_l}) \prod_{l = 1}^{N_d} F_{(i_l, j_l)} |-\rangle. \label{Um}\eeq
We may now directly evaluate the RHS of the above equation by using the bosonized forms (\ref{Feven}) and (\ref{cb}). First,
\beq \prod_{l = 1}^{N_d} F_{(i_l, j_l)} |-\rangle = \frac{1}{2^{N_d/2}}\sum_{\lambda_l = \pm 1} |\{(i_l, j_l)\}^+; \{S^z_{i_l} = \lambda_l, S^z_{j_l} = -\lambda_l\}\rangle \label{sumlambda}.\eeq
Here the index $l$ ranges from $1$ to $N_d$. Further,   $|\{(i_l, j_l)\}^+; \{S^z_{i_l} = \lambda_l, S^z_{j_l} = -\lambda_l\}\rangle$ denotes a state  $|\{S^z, \tau^z\}\rangle$ where $\tau^z = 1$ in the  domains $(i_l, j_l)$, and $\tau^z = -1$ otherwise. $S^z_{m} $ takes the values above for $m = i_l, j_l$, and $S^z_m = 0$ otherwise. After acting on (\ref{sumlambda}) with the $B$ product on the RHS of (\ref{Um}) only one term in the sum over $\lambda_l$ survives: if $n_{i_k} \neq n_{j_k}$ then $\lambda_k = 2 n_{i_k} -1$,  otherwise if $n_{i_k} = n_{j_k}$, $\lambda_k = - (2 n_{i_k} - 1)$. We, thus, focus on just this term. Then after acting with the terms $l = k+1 \ldots N_d$ in the $B$-product in Eq.~(\ref{Um}), up to an overall factor, we get a state with ``$+$ "domains $\{(i_l, j_l)\}$ and $S^z_{i_l} = \lambda_l$, $S^z_{j_l} = -\lambda_l$ for $1 \le l \le k$, and $S^z_{i_l} = 2 n_{i_l} -1, S^z_{j_l} = 2 n_{j_l} - 1$ for $k+1 \le j \le N_d$. 
Thus, once all $B$'s in Eq.~(\ref{Um}) act, up to a phase, we get a state with the same $S^z_i$ as the original state $|\psi\rangle$ and all the $\tau^z$'s flipped. It just remains to compute the phase accumulated during the $B$ action. Let's focus on the $l = k$ term in the $B$-product. If $n_{i_k} \neq n_{j_k}$ we get no phase. If $n_{i_k} = n_{j_k} = 0$  ($n_{i_k} = n_{j_k} = 1$), we are acting with $\sqrt{2} s_{i_1, i_k} c_{i_k}$  ($\sqrt{2} s_{i_1, i_k} c^{\dagger}_{i_k}$). We now use Eq.~(\ref{cb}). The string $\exp\left(\mp\frac{\pi i}{2} \sum_{m = i_1}^{i_k-1} S^z_m\right)$ does not contribute as the sum in the exponent is zero, and we only pick up $(-1)^{N_k+1}$ for $c_{i_k}$ and $(-1)^{N_k}$ for $c^{\dagger}_{i_k}$. Here, $N_k$ is the fermion parity of the state being acted upon. By the above discussion $N_k = \sum_{l = k+1}^{N_d} (n_{i_l} + n_{j_l} - 1)$. Thus, the phase picked up during the action of the $l  =k$ term is
\beq (-1)^{(n_{i_k} + n_{j_k}-1)(N_k + n_{i_k}-1)}\eeq
and the total phase is $(-1)^{\sum_{k = 1}^{N_d} (n_{i_k} + n_{j_k}-1)(N_k + n_{i_k}-1)}$. After some algebra this reduces to $(-1)^{N (N-1)/2}$ where $N = \sum_{l = 1}^{N_d} (n_{i_l} + n_{j_l} -1)$ is the total electron number. This further reduces to $(-1)^{N/2}$ for even $N$, and $-i e^{\pi i N/2}$ for odd $N$. We, thus, recover Eq.~(\ref{US}) of section \ref{sec:unconstr}. 

\subsection{Comparison to Section \ref{sec:unconstr}}
We see that our boundary model in this section matches the ``$S=1$" Hilbert space, operator and symmetry action of section \ref{sec:JW}. (We remind the reader that here we are focusing only on $U(1)$ and unitary $\mathbb{Z}_2$ symmetry, and not time-reversal.) While from the bulk+boundary construction in this section it may appear that one must work in a constrained Hilbert space to describe the boundary, we saw in section \ref{sec:unconstr} that this is not necessary: one can work in an unconstrained local tensor product fermionic Hilbert space. The ``$S = 1$" Hilbert space  is then obtained by adding the term $H_u$, Eq.~(\ref{Huapp}), to the Hamiltonian and taking $u\to \infty$.

\section{Discussion}
In this paper, we've shown that the edge of a QSHI can be mimicked in a 1d lattice model with a local tensor product Hilbert space of finite site dimension. It is interesting what other (super)cohomology phases with continuous symmetry groups share this property. 

Our work also leaves open  the following question. For 2d supercohomology SPT phases with a finite symmetry group $G\times \mathbb{Z}^f_2$ it is possible to construct a finite depth unitary $V$ acting on a Hilbert space of finite site dimension that maps the trivial product state into an SPT ground state, with the property that $V$ commutes with the symmetry (but $V$ is not a product of symmetric local unitaries.)\cite{TylerLukasz, NatAshvin} The existence of such a unitary $V$, in fact, guarantees that the edge can be mimicked without the bulk by employing a non-onsite symmetry.\cite{ElseNayak} It is, thus, interesting, whether such a unitary $V$ exist for the QSHI and, more broadly, for other SPT phases with a continuous symmetry group. Further, if $V$ exists then there is a full commuting projector Hamiltonian realizing the SPT - a property, which may allow one to many-body localize the SPT (assuming that many-body localization exists in dimensions larger than one.) We leave these questions for future work.


\acknowledgements
I would like to thank Robert Jones for a previous collaboration on a related problem and for help with the figures. I am also grateful to Jason Alicea, Dominic Else, Mike Hermele, Itamar Kimchi, T. Senthil, Hassan Shapourian, Jun Ho Son and Ashvin Vishwanath for discussions. This work is supported by the National Science Foundation under grant number DMR-1847861.

\appendix
\section{Bosonization of fermion operators}
\label{app:ferm}
In this appendix we discuss the fermion operators in the treatment of section \ref{sec:unconstr}. This will  be useful in making a comparison to the bulk+edge construction of section \ref{sec:constr}.  We begin in the setting of section \ref{sec:unconstr} by defining 
\beq C_j = \frac12 ((-1)^{g_{j-1,j}(g_{j,j+1}+1)} \gamma_j + i \bar{\gamma}_j).\eeq
$C_j$ is a local fermion operator, obeying $\{C_j, C^{\dagger}_k\} = \delta_{jk}$. Also, 
\beq \tilde{n}_j = C^{\dagger}_j C_j - \frac{1}{4} (1-\tau^z_{j-1,j} \tau^z_{j,j+1}).\eeq
Thus, $[N, C_j] = -C_j$. Also, $[U, C_j] = 0$. 
Thus, $C_j$ has the same quantum numbers as $\psi_R$ in (\ref{psiedge}).

After the JW transformation (\ref{JW}) and unitary rotation (\ref{Rdef}), we obtain
\bea \hat{C}_j &=& \exp(-\pi i \sum_{k = 1}^{j-1} \hat{\tilde{n}}_k) (-i)^{[g_{j-1,j} + g_{j,j+1}]} (-1)^{g_{j-1,j}} \mu^-_j, \quad\quad N - even,\nn\\
 \hat{C}_j &=& \exp(\pi i \sum_{k = 1}^{j-1} \hat{\tilde{n}}_k) (-i)^{[g_{j-1,j} + g_{j,j+1}]}  \mu^-_j, \quad\quad\quad\quad\quad\quad\quad N - odd. \eea
If we restrict $\hat{C}_j$ to the ``spin 1" subspace (ground state subspace of $H_u$ in Eq.~(\ref{HU})), then
\beq \hat{C}_j = -i s_{(1,j)} c_{j}, \eeq
with $c_j$ - the operator in Eq.~(\ref{cb}) obtained via the bulk + boundary construction in section \ref{sec:constr}. Thus, we see that the fermion operators in the two approaches match.

\section{1d Hamiltonian in fermionic variables}
\label{app:fermH}
Here, we write the Hamiltonian (\ref{eq:Hspin1}), (\ref{eq:Hspin2}) in the original fermionic variables of section \ref{sec:symmact}. First, we recall that this Hamiltonian acts in the  ground-state subspace of $H_{u}$, Eq.~(\ref{HU}), which can be written in original variables as
\beq H_u = u \sum_i c^{\dagger}_i c_i (1+ \tau^z_{i-1,i} \tau^z_{i,i+1}). \label{Huapp}\eeq
Next Eqs.~(\ref{eq:Hspin1}), (\ref{eq:Hspin2}) become:
\beq H_{i,i+1} = H^{1}_{i,i+1} + H^{2}_{i,i+1}\eeq
\bea H^1_{i,i+1} &=& - \frac{J}{4} \left((1-c^{\dagger}_i c_i) (1-c^{\dagger}_{i+1} c_{i+1}) - i s_{i,i+1} c^{\dagger}_i c^{\dagger}_{i+1}\right) \tau^x_i (1+\tau^z_{i-1,i} \tau^z_{i,i+1})(1+\tau^z_{i,i+1} \tau^z_{i,i+2}) + h.c., \nn\\
H^2_{i,i+1} &=& - \frac{J}{4} \left((1-c^{\dagger}_i c_i) (1-c^{\dagger}_{i+1} c_{i+1}) + s_{i,i+1} c^{\dagger}_i c^{\dagger}_{i+1}\right) \tau^x_i (1+\tau^z_{i-1,i} \tau^z_{i,i+1})(1-\tau^z_{i,i+1} \tau^z_{i,i+2}) + h.c. \nn\\
\eea
Here, $H^1$ creates/destroys pairs of domain walls, and $H^2$ moves domain walls. It is easy to check that $H^1_{i,i+1}$ and $H^2_{i,i+1}$ commute with $N$ (\ref{Ndef}), $U$ (\ref{Udef}) and $\TNK$ (\ref{TNKdef}).

\section{Anomaly cocycle}
\label{app:restrict}
Here we sketch some steps in the derivation of Eq.~(\ref{w3final}). We begin by writing down the restriction of operators $U^r(\tilde{g})$ to an interval $i \in [1, \ell]$:
\beq {\cal U}^r(\tilde{g}) = (U^r)^{g_1} e^{i g_3 N^r} ({\cal T}^r_{NK})^{g_2}\eeq
where $g_1, g_2 \in \{0,1\}$, $g_3 \in [0, \pi)$  and
\bea U^r &=& \left(\prod_{j=1}^{\ell -1} \tau^x_{j,j+1}\right) \left(\prod_{j = 2}^{\ell -1} e^{-\frac{i \pi}{4}[g_{j-1,j} + g_{j,j+1}]}\right) \left(\gamma^{g_{12}}_1 \left[\prod_{j=2}^{\ell -1} \gamma_j^{g_{j-1,j} + g_{j,j+1}} \right]\gamma_{\ell}^{g_{\ell-1,\ell}}\right), \label{Ur}\\
 N^r &=& \sum_{j=2}^{\ell-1} \tilde{n}_j, \\
 T^r_{NK} &=& \left(\prod_{j=2}^{\ell-1} (-1)^{g_{j,j+1} c^{\dagger}_j c_j}\right) {\cal T}_0.\eea
Terms with smaller $j$ come to the left in the product in square brackets in (\ref{Ur}). 

With the definition above, computing ${\cal U}^r(\tilde{g}) {\cal U}^r(\tilde{h})$ and comparing to ${\cal U}^r(\widetilde{gh})$, we obtain,
\beq L(g,h) = i^{(g_1 h_1 - g_2 h_1 +\lambda^1(g,h)) g_{12}} \gamma^{g_1 h_1}_1 (-1)^{\lambda(g,h) c^{\dagger}_1 c_1},\eeq
with $\lambda(g,h)$ given by Eq.~(\ref{lambda}) and
\beq \lambda^1(g,h) = \langle g_3 + (-1)^{g_2} h_3\rangle \in {\mathbb{Z}}.\eeq
This yields $\sigma(g,h) = g_1 h_1$. Conjugating with ${\cal U}^r(g)$,
\beq ^gL(h,k) = (i (-1)^{g_2})^{(h_1 k_1 - h_2 k_1 +\lambda^1(h,k))(g_{12}-g_1)} \gamma^{h_1 k_1}_1 (-1)^{\lambda(h,k) c^{\dagger}_1 c_1}.\eeq 
Next, computing $w_3(g,h,k)$ from (\ref{w3def}) we find
\beq w_3(g,h,k) = i^{g_1 (h_1 k_1 + (-1)^{g_2} h_2 k_1 + (-1)^{g_2} \lambda^1(h,k))} (-1)^{g_2 h_1 k_1 + g_1 h_1 h_2 k_1 + (g_1 + h_1) \lambda^1(g,h) k_1 + g_1 (h_1 + k_1) \lambda^1(h,k)}. \eeq
Performing a gauge transformation, $w'_3 = w_3  (d_T \gamma)$, with
\beq \gamma(g,h) = \exp\left(\frac{\pi i}{4}(g_2 h_1 + 2 g_1 g_2 h_1 + 2 [g_1+h_1] \lambda^1(g,h))\right),\eeq 
we obtain 
\beq w_3(g,h,k) = i^{g_1 h_1 k_1 + \lambda^1(g,h) k_1},\eeq
in agreement with Eq.~(\ref{w3final}). 

\section{Boundary of a general supercohomology 2d SPT with a finite symmetry group $G_f$.}
\label{app:Ugen}
In this appendix we construct the boundary symmetry action for any 2d supercohomology fermion SPT with a finite symmetry group $G_f$. As explained in section \ref{sec:cocmain},  it is useful to form the group $G_b = G_f/\mathbb{Z}^f_2$ and find the associated cocycle $\lambda \in H^2(G_b, \Z_2)$, (\ref{lambdadef}). Then the non-onsite boundary symmetry is characterized by two pieces of algebraic data: $\sigma \in H^2(G_b, \Z_2)$ and $w_3 \in C^3(G_b, U(1))$ satisfying Eq.~(\ref{dTw3}) and modulo gauge transformations (\ref{gaugecoc}). 

We let the boundary Hilbert space be a 1d chain with complex fermions $c_i$ living on sites $i = 1 \ldots L$, and group elements $g_{i,i+1} \in G_b$ living on links. As before, we split $c_i$ into Majorana fermions (\ref{cgamma}). For $g \in G_b$, we let
\bea {\cal U}(\tilde{g}) = \prod_{i = 1}^{L} U^0_{i,i+1}(g) \prod_{i =1}^{L} \tilde{w}^{-1}_3(g, g_{i-1,i}, g_{i-1,i}^{-1}g_{i,i+1}) \prod_{i=1}^{L} \gamma_i^{\sigma(g,g_{i-1,i}) + \sigma(g,g_{i,i+1})} \prod_{i = 1}^{L} (-1)^{\lambda(g, g_{i,i+1}) c^{\dagger}_i c_i} \,\,{\cal T}^{\rho(g)}_0 \,.\nn\\
\label{Ugenconst}\eea
Here $U^0_{i,i+1}(g)$ acts only on link $(i,i+1)$: $U^0_{i,i+1}(g)|g_{i,i+1}\rangle = | g\cdot g_{i,i+1}  \rangle$. The terms with smaller $i$ come to the left in the $\gamma$ product. $\rho(g) = 1$ if $g$ is time-reversing and $\rho(g) = 0$ otherwise. The factor $\tilde{w}_3$ is defined to be
\beq \tilde{w}_3(g,h,k) = w_3(g,h,k) (-1)^{(\sigma(g,hk) + \lambda(g,hk)) \sigma(h,k)}.\eeq
One can check that ${\cal U}(\tilde{g})$, indeed, satisfies the algebra (\ref{lambdadef}). 

We next check that ${\cal U}(\tilde{g})$ defined above is, indeed, characterized by the data $\sigma$ and $w_3$ that went as input into (\ref{Ugenconst}). We form the restriction of ${\cal U}(\tilde{g})$ to the interval $i \in [1,\ell]$ as follows:
\bea {\cal U}^r(\tilde{g}) &=&  \prod_{i = 1}^{\ell-1} U^0_{i,i+1}(g) \prod_{i =2}^{\ell-1} \tilde{w}^{-1}_3(g, g_{i-1,i}, g_{i-1,i}^{-1}g_{i,i+1}) \nn\\&\times& \gamma^{\sigma(g,g_{12})}_1 \prod_{i=2}^{\ell-1} \gamma_i^{\sigma(g,g_{i-1,i}) + \sigma(g,g_{i,i+1})} \gamma^{\sigma(g,g_{\ell-1,\ell})}_\ell \,\, \prod_{i = 2}^{\ell-1} (-1)^{\lambda(g, g_{i,i+1}) c^{\dagger}_i c_i} \,\,{\cal T}^{\rho(g)}_0 \,. \eea
Computing $ {\cal U}^r(\tilde{g})  {\cal U}^r(\tilde{h})$ we extract $L(g,h)$ in Eq.~(\ref{Lghdef}): 
\beq L(g,h)  =  \gamma_1^{\sigma(g,h)} (-1)^{\lambda(g,h) c^{\dagger}_1 c_1}  w_3(g, h, (gh)^{-1} g_{12}) (-1)^{\lambda(g, h) \sigma(gh, (gh)^{-1} g_{12}) + \sigma(g, g^{-1} g_{12}) \sigma(h, (gh)^{-1} g_{12})}.\eeq
From this we learn that $\sigma$ is, indeed, the two-cocycle characterizing the non-onsite symmetry. Finally, from Eq.~(\ref{w3def}) we find that $w_3$ is, indeed, the three-cochain characterizing the non-onsite symmetry.

\bibliography{Insulator}

\end{document}